\documentclass{ws-procs975x65} 

\usepackage{float}
\usepackage{braket}
\usepackage{color,xcolor}

\begin{document}

\title{LARGE SCALE STRUCTURES: FROM INFLATION TO TODAY: A BRIEF REPORT}

\author{Lina Castiblanco ; Radouane Gannouji and Cl\'ement Stahl}

\address{Instituto de F\'{\i}sica, Pontificia Universidad Cat\'{o}lica de Valpara\'{\i}so, Casilla 4950, Valpara\'{\i}so, Chile\\
lina.castiblanco.t@mail.pucv.cl ; radouane.gannouji@pucv.cl ; clement.stahl@pucv.cl}

\begin{abstract}
We briefly review some selected topics gravitating around large scale structures. We derive from inflation the evolution of dark matter perturbations. The stress is put on the non-linear regime of structures formation, with a particular emphasis on relativistic effects, the Effective Field Theory approach, the role of dark energy and the possibility of inhomogeneous universes.
\end{abstract}



\bodymatter 

\vskip1cm
\section{Introduction and spirit of the review:}

This story started with the theory of general relativity (GR) that provided the laws guiding the evolution of the universe and the structures. Few years later, in 1922, Friedmann found a dynamical solution of a homogeneous and isotropic universe. These equations were rederived and confronted to the observations by Lema\^itre in 1927 \cite{Lemaitre:1927zz}. In 1929, Hubble \cite{Hubble:1929ig} using his measure of distances and Slipher’s redshift data \cite{Slipher} rediscovered the results found by Lema\^itre \cite{Lemaitre:1927zz} which is now called the Hubble law. The universe is expanding and should have an early time which is hot and dense. These ideas converged to the so-called Big Bang model. It describes an expanding universe composed principally by radiation, later by matter and finally dominated by a cosmological constant, this is the $\Lambda$CDM model (Lambda Cold Dark Matter). It became the standard and very successful model in cosmology. In this review, we are principally interested in the large scale structures, or how the structure forms over the evolution of the universe. We will describe how from initial conditions defined during inflation we can arrive to structures at late time.

Those structures are understood as little perturbations in an otherwise expanding universe. The dynamics of these little perturbations is understood in the so-called linear universe when it was young enough. Sections \ref{sec:inflation} and \ref{sec:linear} review them in some details. In section \ref{sec:non-lin}, we will present how these perturbations evolve in the non-linear regime, where astrophysical structures start to form. This regime requires much more advanced mathematical techniques to better model this complicated non-linear situation. The focus of section \ref{sec:non-lin} will be standard perturbation theory. In section \ref{sec:relat}-\ref{sec:EFTofLSS}, we present three extensions of the formalism: relativistic corrections, dark energy and a model for the small scales physics with effective field theories. In section \ref{sec:Inhom}, we present a totally alternative point of view on the evolution of perturbations: relaxing the hypothesis of a homogeneous background. This section is also there to remind the reader that any cosmological observation is done by assuming a cosmological model at the first place. Finally in section \ref{sec:concl}, we conclude and give perspective for the future of the field. 

Already excellent reviews\cite{Ma:1995ey,Bernardeau:2001qr,Biagetti:2019bnp,Porto:2016pyg} and textbooks\cite{Dodelson:2003ft,Liddle:2000cg,Baumann:2009ds,Peter:2013avv,Baumann:2014nda,Piattella:2018hvi} covered most of the material that we wish to discuss here. The originality of our approach is to cover structures formation from the inflation to the non-linear dark matter power spectrum. We also want to offer an introduction to the current status in 2019 of the field. Open problems are discussed elsewhere.\cite{Coley:2019yov}

\paragraph{Notations} A dot denote a derivative with respect to cosmic time $t$  and a prime is the conformal time derivative: $'\equiv d/d\eta$. The Laplacian is written $\Delta$.
We denote $\delta_D$ as the Dirac delta distribution. Our Fourier convention is:
\begin{equation}
      f(\bm{x})=\int\frac{{\rm d}^3\textbf{k}}{(2\pi)^3}f(\bm{k})e^{i\bm{k}.\bm{x}},\\
\end{equation}
and we sometimes use the shorthand for Fourier space integration:

\begin{align}
    \int_{\bm{ k}_1,\bm{k}_2}=\int\frac{{\rm d}^3\textbf{k}_1}{(2 \pi)^3}\frac{{\rm d}^3\textbf{k}_2}{(2 \pi)^3}\delta_{D}(\bm{k}_1+\bm{k}_2-\bm{k})
\end{align}

$\delta(t,\bm{x})$ is the density contrast, described by a stochastic variable. Observables will be related to correlations of $\delta(t,\bm{x})$.
Due to translational and rotational invariance, the information of the two and three point correlation function can be encapsulated in the power spectrum and the bispectrum:
\begin{align}
    & \langle \delta(t,\bm{k}_1) \delta(t,\bm{k}_2)\rangle= (2\pi)^3 \delta_D(\bm{k}_1+\bm{k}_2) P(t,k_1)\,, \label{eq:2ptcorr}\\
    & \langle \delta(t,\bm{k}_1) \delta(t,\bm{k}_2)\delta(t,\bm{k}_3)\rangle=(2\pi)^3 \delta_D(\bm{k}_1+\bm{k}_2+\bm{k}_3) B(t,k_1, k_2,k_3)\,. \label{eq:3ptcorr}
\end{align}
The bracket denotes an ensemble average which will agree with statistical average under ergodicity assumption.
Similarly $P_{\mathcal{R}}$ and $P_{\psi}$ are defined by taking correlators of $\mathcal{R}(t,\textbf{k})$ and $\psi(t,\textbf{k})$.

\section{Inflation}
\label{sec:inflation}
Inflation describes an early acceleration of the expansion of the universe that solves some of the fundamental questions in cosmology such as the horizon, the flatness or the magnetic monopole problem. For that, a scalar field known as the inflaton, produces a quasi-de-Sitter acceleration. This simple inflationary paradigm proposes an elegant solution to the previous mentioned problems but it became much more interesting by suggesting a mechanism to understand the observed structures in the universe. From quantum fluctuations at microscopic scales which are magnified to cosmic scales, the seeds for the growth of structures in the universe emerge. At the end of the accelerating phase, the inflaton supposedly decays into ordinary matter during reheating, and this process could be very difficult to calculate. One strategy is to work with a variable $\mathcal{R}(t,\bm{x})$ which is constant on superhorizon scales and for adiabatic perturbations. It will have again a dynamics once the modes reenter the horizon, which means after reheating. Therefore, we will have a method to relate the dynamics of our fields from the radiation epoch to today starting from initial conditions defined during inflation while passing the complexity of reheating without any effects. The difficulty would be to relate this new variable $\mathcal{R}(t,\bm{x})$ to an observable today. We will first, introduce this new variable and see its dynamics before relating it to the gravitational potential $\phi(t,\bm{x})$ which will be our link to an observable. \\
Considering a FLRW (Friedmann-Lema\^itre-Robertson-Walker) metric with scalar perturbations in the Newtonian gauge\footnote{see section \ref{sec:relat} for a discussion on the different gauges.}
\begin{align}
\label{eq:metricLIN}
{\rm d}s^2=-(1+2\phi){\rm d}t^2+a^2(t)(1-2\psi)\delta_{ij}{\rm d}x^i{\rm d}x^j.
\end{align}
We define a gauge invariant quantity $\mathcal{R} \equiv \psi-H \delta q/(\rho+P)$ where $\delta q$ is the scalar part of the 3-momentum density which is related to the velocity $v_i$ by $T_i^0=\partial_i \delta q=(\rho+P)a\partial_i v$, $\rho$ and $P$ are the energy density and pressure respectively and $H=\frac{\dot{a}}{a}$. Using line element~\eqref{eq:metricLIN}, we find for Einstein equations $G_i^0$
\begin{align}
\dot\psi+H\psi=-4\pi G \delta q,
\end{align}
where we have used $\phi=\psi$ that can be obtained from $G_i^j$ in the absence of anisotropic stress. Therefore, the comoving curvature perturbation may be written as
\begin{align}
\label{eq:R}
\mathcal{R}=\psi-\frac{H\delta q}{\rho+P}=\psi+\frac{H}{4\pi G(\rho+P)}(\dot\psi+H\psi)=\psi+\frac{2}{3}\frac{\dot\psi+H\psi}{H(1+w)},
\end{align}
where we have used the Friedmann equation and the notation $w=P/\rho$. The Einstein equation $G_i^i$ gives
\begin{align}
\label{eq:Psi}
\ddot\psi+4H\dot\psi+(3H^2+2\dot H)\psi=4\pi G \delta P,
\end{align}
where $\delta P$ is the pressure perturbation. For adiabatic perturbations we can relate the pressure to the density $\delta P=c_s^2 \delta\rho$ where $c_s^2$ is the adiabatic sound speed and $\delta\rho$ is related to the gravitational potential through the $(0,0)$ component of the Einstein equations which in Fourier space is given by
\begin{align}
\label{eq:PoissonGR}
3 H \dot \psi+3 H^2\psi+\frac{k^2}{a^2}\psi=-4\pi G \delta\rho.
\end{align}
Using this expression, we get for Eq.~(\ref{eq:Psi})
\begin{align}
\ddot\psi+\Bigl[4H+3Hc_s^2\Bigr]\dot\psi+\Bigl[3H^2(1+c_s^2)+2\dot H\Bigr]\psi+c_s^2 \frac{k^2}{a^2}\psi=0.
\end{align}
Finally, writing $c_s^2=\dot P/\dot \rho=w-\dot w/(3H(1+w))$, we have
\begin{align}
\label{eq:ddotpsi}
\ddot\psi+\Bigl(4H+3Hw-\frac{\dot w}{1+w}\Bigr)\dot\psi-H\frac{\dot w}{1+w}\psi+c_s^2 \frac{k^2}{a^2}\psi=0.
\end{align}
Taking the derivative of Eq.~(\ref{eq:R}) and using Eq.~(\ref{eq:ddotpsi}), we get
\begin{align}
\frac{3H(1+w)}{2c_s^2}\mathcal{\dot R}=-\frac{k^2}{a^2}\psi.
\end{align}
It is clear that on superhorizon scales, where $k\ll a H$, the right hand side becomes negligible and thus $\mathcal{R}(t,\bm{k})$ is constant in time. This result tells us that when a wavelength generated during inflation evolves on superhorizon scales, $\mathcal{R}(t,\bm{k})$ is constant in time and becomes dynamical only when the given wavelength re-enters the horizon, therefore we can skip all possible complicated physics after inflation until radiation era. As we will see later, we can connect the observable which is the correlation function of matter to the gravitational potential $\psi(t,\bm{k})$, which can be connected to the comoving curvature perturbation. In fact, considering that our wavelength re-enters the horizon during an era when $a(t)\sim t^{2/(3(1+w))}$, where $w$ is constant, Eq.~(\ref{eq:ddotpsi}) gives $\psi(t,\bm{k})=A+B t^{-\frac{5+3w}{3+3w}}$ which corresponds to a constant and a decaying mode. Plugging this expression in Eq.~(\ref{eq:R}), we get
\begin{align}
\label{eq:psiR}
\psi\simeq \frac{3(1+w)}{5+3w}\mathcal{R}.
\end{align}
Considering that most of interesting modes re-enter during radiation era $(w=1/3)$, we get $\psi=2\mathcal{R}/3$. Therefore the power spectrum of the gravitational perturbations during radiation-domination reads
\begin{align}
P_{\psi}(t,\bm{k})=\frac{4}{9}P_{\mathcal{R}}(t,\bm{k}).
\end{align}
We will see later, thanks to the Boltzmann equations, how to find the power spectrum at late time from the power spectrum during radiation era. But for now, we would like to find the $P_{\mathcal{R}}(k)$ which is much simpler to obtain because it is the power spectrum generated during inflation ($\mathcal{R}(t,\bm{k})$ remains constant in time during reheating). To calculate $P_{\mathcal{R}}(k)$, we just need to follow some standard calculations which are described in many textbooks and lectures\cite{Liddle:2000cg,Dodelson:2003ft,Baumann:2009ds,Maldacena:2002vr} and therefore we will not reproduce them here. The final result reads as
\begin{align}
k^3 P_{\mathcal{R}}(k)=2\pi^2A_s\Bigl(\frac{k}{k_*}\Bigr)^{n_s-1},
\end{align}
where $n_s$ is the scalar spectral index which encodes the scale-dependence of the power spectrum, $k_*$ is an arbitrary reference or pivot scale and $A_s$ is the scalar power spectrum amplitude.
Surprisingly, this parametrization of the power spectrum can be evolved until the emission of the Cosmic Microwave Background (CMB) and CMB experiments were able to measure $n_s$ and $A_s$ with a great deal of precision.\cite{Aghanim:2018eyx} For the rest of this review, the focus will be to evolve those perturbations until the recent time and evaluate the impact of those parameters on the late time distribution of dark matter.
\section{From inflation to late time universe: the evolution of the linear power spectrum}
\label{sec:linear}
In this section, we will obtain the linear power spectrum at any time of the evolution of the universe. Having in mind the power spectrum predicted by inflation, we need to calculate its evolution in time until today when it is observed by our telescopes. This evolution can be decomposed in two parts. There is a linear evolution which dominates in the early universe and at large scales. It will be the objective of this section to calculate the evolution of the power spectrum at linear order. For more details, see these interesting books\cite{Dodelson:2003ft,Piattella:2018hvi} and review\cite{Ma:1995ey}. In section 4, we will introduce non-linear corrections which occur at small scales. 

\subsection{Boltzmann equation}

We need to develop some useful machinery to describe how the entire cosmological density field and all other physical quantities evolves as function of time. We could take only the equations of GR but being a theory describing gravity, it doesn't contain all possible interactions between particles which will be encoded in the Boltzmann equation:
\begin{align}
\frac{{\rm d}f}{{\rm d}t}=C[f],
\end{align}
where $f(t,x^i,p^i)$ is the distribution function and $C[f]$ represents all possible interactions. The one-particle distribution is a function of the coordinates of our phase space: the time, the particle position and momentum. Working in a covariant way, we will abandon the time for an affine parameter $\lambda$ and obtain
\begin{align}
\frac{{\rm d}f}{{\rm d}\lambda}=C[f].
\end{align}
We could yet write our distribution function as $f(x^\mu,p^\mu)\equiv f(t,x^i,p^\mu)$. But doing that, we would have $f$ as a function of 8 variables instead of the 7 ($t,x^i,p^i$) that we had previously. But in fact, $p^\mu$ has a constraint $p^\mu p_\mu=-m^2$, which permits us to eliminate $p^0$ from the variables. Considering that our metric will be diagonal, even at level of perturbation, we have $g_{00}(p^0)^2=-m^2-p^2$ where $p^2=g_{ij}p^i p^j$. Also we can replace $p^i$ for its norm $p$ and the direction vector $\hat p^i$ which satisfies $\delta_{ij}\hat p^i \hat p^j=1$. Finally we can always replace $p$ for the energy defined as $E=\sqrt{p^2+m^2}$. Therefore we have $f\equiv f(t,x^i,E,\hat p^i)$. Expanding the total derivative, we have
\begin{align}
\label{eq:df}
\frac{{\rm d}f}{{\rm d}\lambda}=\frac{{\rm d}t}{{\rm d}\lambda}\frac{\partial f}{\partial t}+\frac{{\rm d}x^i}{{\rm d}\lambda}\frac{\partial f}{\partial x^i}+\frac{{\rm d}E}{{\rm d}\lambda}\frac{\partial f}{\partial E}+\frac{{\rm d}\hat p^i}{{\rm d}\lambda}\frac{\partial f}{\partial \hat p^i}.
\end{align}
Considering a FLRW Universe as our background metric, any dependence on the direction $\hat p^i$ can be only of perturbations order and therefore the last term of Eq.~(\ref{eq:df}) is second order and can be neglected. We have $\frac{{\rm d}t}{{\rm d}\lambda}=p^0$ and $\frac{{\rm d}x^i}{{\rm d}\lambda}=p^i$. Using Eq.~(\ref{eq:metricLIN}), we get $p^0=E/\sqrt{1+2\phi}\simeq E(1-\phi)$, and we can write $p^i=\alpha \hat p^i$ where $\alpha$ is a normalization factor. We have $p^2=a^2(1-2\psi) \alpha^2 \hat p^i \hat p_i=a^2(1-2\psi) \alpha^2$, which implies $\alpha=p a^{-1} (1-2\psi)^{-1/2}$ and therefore $p^i/p^0\simeq p \hat p^i/aE$. In order to obtain $\frac{{\rm d}E}{{\rm d}\lambda}$ we use the geodesic equation
\begin{align}
\frac{{\rm d}p^0}{{\rm d}\lambda}+\Gamma_{\mu \nu}^0p^\mu p^\nu=0.
\end{align}
Using that $\frac{{\rm d}p^0}{{\rm d}\lambda}=p^0\frac{{\rm d}p^0}{{\rm d}t}$, the perturbed metric for $\Gamma_{\mu \nu}^0$ and $p^0=E(1-\phi)$, we have at first order
\begin{align}
\frac{{\rm d}E}{{\rm d}t}+\frac{p}{a}\hat p^i\partial_i \phi+\frac{p^2}{E}(H-\dot \psi)=0,
\end{align}
which gives the Boltzmann equation at first order
\begin{align}
\frac{\partial f}{\partial t}+\frac{p\hat p^i}{a E}\partial_i f-p\frac{\partial f}{\partial E}\Bigl(\frac{\hat p^i}{a}\partial_i \phi+\frac{H p}{E}-\frac{p}{E}\dot \psi\Bigr)\simeq \frac{1}{E}C[f].
\end{align}
For the interaction term, we will consider a generic interaction with particles $(a,b,c,d)$ and momentum $(\mathbf{p},\mathbf{q},\mathbf{p}',\mathbf{q}')$ of the following form $a(\mathbf{p})+b(\mathbf{q})\rightarrow c(\mathbf{p}')+d(\pmb{q}')$. We have for the distribution function of the particle $a$
\begin{align}
C[f_a(\mathbf{p})]&=\frac{1}{p}\int\frac{{\rm d}^3 \mathbf{q}}{(2\pi)^3 2 E_b(q)} \int\frac{{\rm d}^3 \mathbf{p}'}{(2\pi)^3 2 E_c(p')} \int\frac{{\rm d}^3 \mathbf{q}'}{(2\pi)^3 2 E_d(q')}|\mathcal{M}|^2\delta_D(\mathbf{p}+\mathbf{q}-\mathbf{p}'-\mathbf{q}')\nonumber\\
&\delta_D(E_a(\mathbf{p})+E_b(\mathbf{q})-E_c(\mathbf{p}')-E_d(\mathbf{q}'))\Bigl[f_c(\mathbf{p}')f_d(\mathbf{q}')(1\pm f_a(\mathbf{p}))(1\pm f_b(\mathbf{q}))\nonumber\\
&-f_a(\mathbf{p})f_b(\mathbf{q})(1\pm f_c(\mathbf{p}'))(1\pm f_d(\mathbf{q}'))\Bigr],
\end{align}
where $|\mathcal{M}|^2$ is the amplitude which encodes the information about the interaction (that is assumed symmetric). As usual in textbooks\cite{Peter:2013avv} of cosmology, the particle degeneracy $g_s$ of each particle is incorporated in the distribution functions, $f_i$. Finally, because of the Pauli exclusion principle, it is easier to produce a boson than a fermion which is incorporated in this formula through the coefficients $1\pm f$ which are known as Bose enhancement and Pauli blocking.

\subsection{Cold dark matter}
For dark matter, we don't have any interaction term and therefore the Boltzmann equation is
\begin{align}
\label{eq:BoltzmannDM}
\frac{\partial f_{dm}}{\partial t}+\frac{p\hat p^i}{a E}\partial_i f_{dm}-p\frac{\partial f_{dm}}{\partial E}\Bigl(\frac{\hat p^i}{a}\partial_i \phi+\frac{H p}{E}-\frac{p}{E}\dot \psi\Bigr)=0.
\end{align}
To obtain our equations in a more standard form, we need to take moments of this equation. Considering the zeroth moment, which means integrating over $\rm{d}^3\mathbf{p}$, we obtain
\begin{align}
\label{eq:DM}
&\frac{\partial }{\partial t}\int\frac{{\rm d}^3 \mathbf{p}}{(2\pi)^3}f_{dm}+
\frac{1}{a}\frac{\partial }{\partial x^i}\int\frac{{\rm d}^3 \mathbf{p}}{(2\pi)^3}f_{dm}\frac{p\hat p^i}{E}
-\Bigl(H-\dot\psi\Bigr)\int\frac{{\rm d}^3 \mathbf{p}}{(2\pi)^3}\frac{\partial f_{dm}}{\partial E}\frac{p^2}{E}\nonumber\\
&-\frac{1}{a}\frac{\partial \phi}{\partial x^i}\int\frac{{\rm d}^3 \mathbf{p}}{(2\pi)^3}\frac{\partial f_{dm}}{\partial E}p \hat p^i=0
\end{align}
and because we have the dark matter density and velocity defined as 
\begin{align}
& n_{dm}(t,\bm{x})=\int\frac{{\rm d}^3 \mathbf{p}}{(2\pi)^3}f_{dm}(t,\bm{x},E,\hat p^i),\\
& n_{dm}v^i(t,\bm{x})=\int\frac{{\rm d}^3 \mathbf{p}}{(2\pi)^3}\frac{p\hat p^i}{E}f_{dm}(t,\bm{x},E,\hat p^i),
\end{align}
we obtain from Eq.~(\ref{eq:DM}) at first order
\begin{align}
\label{eq:zeromoment}
\dot n_{dm}+\frac{1}{a}\partial_i (n_{dm}v^i)+3 n_{dm}(H-\dot\psi)=0.
\end{align}
Considering the deviation from the homogeneous value 
\begin{align}
    n_{dm}(t,\bm{x})=\bar n_{dm}(t)\left[1+\delta(t,\bm{x})\right]\,,
\end{align}
we get at first order
\begin{align}
\dot\delta+\frac{\partial_i v^i}{a}-3\dot\psi=0.
\end{align}
Working in Fourier space, 
we have
\begin{align}
\dot\delta+\frac{ik_i v^i}{a}-3\dot\psi=0,
\end{align}
or if the fluid is irrotational $v^i=k^i v/k$ (that we will always assume)
\begin{align}
\label{eq:system1}
\dot\delta+\frac{ik v}{a}-3\dot\psi=0.
\end{align}
Notice that this equation could also be derived from $\nabla_\mu T^\mu_{~\nu}=0$ because of the absence of interactions. Considering now the first moment, i.e. multiplying equation (\ref{eq:BoltzmannDM}) by $p\hat p^i/E$ and integrating over ${\rm d}^3\mathbf{p}$, we get\footnote{Being in a euclidean space, we have $\delta_{ij}\partial^j=\partial_i$ and therefore we will write indifferently lower and upper indices when necessary.}
\begin{align}
\dot v^i+H v^i+\frac{\partial_i \phi}{a}=0,
\end{align}
or for the irrotational fluid in the Fourier space
\begin{align}
\label{eq:system2}
\dot v+H v+\frac{i k \phi}{a}=0,
\end{align}
where we have neglected higher order terms in $p/E$ because $p/m \sim v\ll 1$. This condition is related to the fact that we consider cold dark matter, for which the velocity is small. In that case, higher order terms in $p/E$ are negligible and therefore we do not need to look for higher moments.
\subsection{Baryons}
Another component of the universe, which is important are the so-called baryons or more exactly the electrons and protons. In that case the over-densities of electrons and protons are equal $\delta_e=\delta_p\equiv \delta_b$ as well as their velocities $v_e=v_p\equiv v_b$ because of the tight coupling between them through Coulomb scattering $(e+p\rightarrow e+p)$. The derivation of the equations for baryons will be very similar to cold dark matter, using moments, but we will need to include interactions. We need to consider Coulomb scattering and Compton scattering $(e+\gamma\rightarrow e+\gamma)$.  This last interaction will be the only contribution to the first moments. For that, we will need to introduce the temperature of the photons and its perturbation $\theta(t,\bm{x})=\delta T(t,\bm{x})/T(t)$, also the amplitude $|\mathcal{M}|^2=8\pi\sigma_T m_e^2$ where $\sigma_T$ is the Thomson cross-section and $m_e$ the mass of electrons. Finally, we define the optical depth $\dot \tau=-n_e \sigma_T$ where $n_e$ is the number of free electrons. We get for the zeroth moment
\begin{align}
\label{eq:system3}
\dot\delta_b+\frac{ik v_b}{a}-3\dot\psi=0.
\end{align}
Notice that interactions do not appear in this equation because of conservation number of electrons during the scattering process. The first moment contains interaction and gives
\begin{align}
\label{eq:system4}
\dot v_b+H v_b+\frac{ik \phi}{a}=\dot \tau R(3i\theta_1+v_b),
\end{align}
where $\theta_1$ is the first moment of $\theta$ defined as $\theta_1 \equiv i/2\int_{-1}^1\mu \theta(\mu){\rm d}\mu$ and $R \equiv \frac{4\rho_\gamma}{3\rho_b}=\frac{4\Omega_{r0}}{3a\Omega_{b0}}$, $\rho_\gamma$ is the energy density of radiation while $\rho_b$ is the energy density of baryons. We have introduced the first moment of a function. Generically we can define the expansion in multipoles as
\begin{align}
\label{eq:thetaL}
\theta_l &=\frac{i^l}{2}\int_{-1}^1{\rm d}\mu P_l(\mu)\theta(\mu),\\
\label{eq:thetaL2}
\theta(\mu) &=\sum_{l=0}^\infty \frac{2l+1}{i^l}\theta_l P_l(\mu).
\end{align}
where we defined $\mu=k^i \hat p_i/k$ the cosine of the angle between the wavenumber $\bm{k}$ and the direction $\bm{p}$, and $P_l$ are Legendre polynomials of degree $l$.

\subsection{Photons}
As we have seen previously, because of the coupling between baryons and photons, we have $\theta_1$ which therefore needs an equation of evolution. For that we will consider the Boltzmann equation for photons, so we have $E=p$. The strategy will be different than previous cases. In fact, for photons, we know the form of the photon distribution function $f(t,x^i,E,\hat{p}^i)$ at the background level, $f^{(0)}(t,p)$, the Bose-Einstein distribution $f^{(0)}(t,p)=(e^{p/T(t)}-1)^{-1}$. Considering now a perturbation around it, we have $f(t,x^i,E,\hat{p}^i)=\left[\exp \left({\frac{p}{T(1+\theta)}}\right)-1\right]^{-1}\simeq f^{(0)}(t,p)-p\theta \partial f^{(0)}/\partial p$. 

The Compton scattering will not have any influence at the zero-order and the Boltzmann equation gives simply $dT/T=-da/a$ which is $T(t)\propto 1/a(t)$. Considering now the equation at first order, but not the first moment because we do not integrate over ${\rm d}^3 \mathbf{p}$, and also adding the interaction term, we have 
\begin{align}
\dot\theta+\frac{1}{a}\hat p^i\partial_i \theta-\dot \psi+\frac{1}{a}\hat p^i\partial_i \phi=-\dot \tau (\theta_0-\theta+\hat p_i.v_b^i),
\end{align}
which gives in Fourier space,  
\begin{align}
\label{eq:system5}
\dot\theta-\dot \psi+\frac{i k \mu}{a}(\theta+\phi)=-\dot \tau (\theta_0-\theta+\mu v_b),
\end{align}
where we assumed an irrotational fluid, so $v^i=k^i v/k$.

Neutrinos will follow exactly the same equation but without Compton scattering, that we will describe by the variable $\mathcal{N}(t,\bm{x})$. 
\begin{align}
\label{eq:system5bis}
 \mathcal{\dot N}-\dot \psi+\frac{i k \mu}{a}(\mathcal{N}+\phi)=0.
\end{align}
We could also add an equation for the polarization but we will neglect it.

\subsection{Einstein equations for scalar perturbations}
To close our previous system of equations, we need 2 additional equations, which will describe the dynamics of the gravitational potentials $(\phi,\psi)$. From the metric Eq.~(\ref{eq:metricLIN}), we have at first order of perturbations
\begin{align}
&\delta G_0^0 =6H(\dot\psi+H\phi)-2\frac{\Delta\psi}{a^2},\\
\label{eq:deuxieme}
&\delta G_i^j-\frac{1}{3}\delta_i^j \delta G_k^k =\frac{1}{a^2}(\partial_i\partial^j-\frac{1}{3}\delta_i^j\Delta)(\psi-\phi),
\end{align}
which is equal to the energy momentum tensor. The first equation should be equal to $-8\pi G\sum_i \delta \rho_i$, where $i$ represents the different components in the universe. In the Fourier space, we get
\begin{align}
-3H(\dot\psi+H\phi)-\frac{k^2}{a^2}\psi=4\pi G(\rho_{dm}\delta+\rho_b\delta_b+\rho_\gamma\delta_\gamma+\rho_\nu\delta_\nu).
\end{align}
But because we want to use the variables previously introduced, we have
\begin{align}
\delta \rho_i = g_i \int\frac{{\rm d}^3\mathbf{p}}{(2\pi)^3}E_i(p) \delta f_i.
\end{align}
For photons, we have a spin degeneracy $g_\gamma=2$, $\delta f_\gamma=-p\partial f/\partial p$ and $E=p$:
\begin{align}
\delta \rho_\gamma &=-2 \int\frac{{\rm d}^3\mathbf{p}}{(2\pi)^3} p^2 \frac{\partial f_\gamma}{\partial p}\theta=-\frac{1}{4\pi^3} \int p^4 \frac{\partial f_\gamma}{\partial p}{\rm d}p \int{\rm d}\Omega ~\theta=-\frac{\theta_0}{\pi^2} \int p^4 \frac{\partial f_\gamma}{\partial p}{\rm d}p \nonumber\\
&=4\frac{\theta_0}{\pi^2} \int p^3 f_\gamma {\rm d}p=4\rho_\gamma \theta_0,
\end{align}
where we used that 
\begin{align}
\rho_\gamma=2\int \frac{{\rm d}^3\mathbf{p}}{(2\pi)^3}p f_\gamma=\frac{1}{\pi^2}\int {\rm d}p p^3 f_\gamma.
\end{align}
Same calculations can be performed for neutrinos, which gives
\begin{align}
\label{eq:system6}
-3H(\dot\psi+H\phi)-\frac{k^2}{a^2}\psi=4\pi G(\rho_{dm}\delta+\rho_b\delta_b+4\rho_\gamma\theta_0+4\rho_\nu\mathcal{N}_0).
\end{align}
For Eq.~(\ref{eq:deuxieme}), we have in Fourier space
\begin{align}
k^2(\hat k_i \hat k_j-\frac{1}{3}\delta_{ij} )(\phi-\psi)=8\pi G a^2 \pi_{ij},
\end{align}
where $\pi_i^j$ is the anisotropic stress. Multiplying this expression by $\hat k^i\hat k^j$, we get
\begin{align}
\label{eq:k2}
k^2(\phi-\psi)=12\pi G a^2 \pi_{ij}\hat k^i \hat k^j.
\end{align}
But considering 
\begin{align}
\delta T_\mu^\nu=\sum_i g_i \int \frac{{\rm d}^3\mathbf{p}}{(2\pi)^3} \frac{p_\mu p^\nu}{p^0}\delta f_i
\end{align}
and 
\begin{align}
\pi_{i}^j=\delta_i^j-\delta_i^j \delta T_k^k/3=\int \frac{{\rm d}^3\mathbf{p}}{(2\pi)^3} \frac{1}{p^0} (p_i p^j-\frac{1}{3}\delta_i^j p^2)\delta f,
\end{align}
we have for photons
\begin{align}
\pi_{ij}\hat k^i\hat k^j=-2 \int \frac{{\rm d}^3\mathbf{p}}{(2\pi)^3} p^2(\mu^2-\frac{1}{3})\frac{\partial f}{\partial p}\theta.
\end{align}
We identify the second Legendre polynomial $P_2(\mu)=(3 \mu^2-1)/2$, and integrating over the angles ($\int {\rm d}\Omega=2\pi\int_{-1}^1 {\rm d}\mu$), we obtain
\begin{align}
\pi_{ij}\hat k^i\hat k^j &=-\frac{1}{3\pi^2}\int {\rm d}p p^4\frac{\partial f_\gamma}{\partial p}\int_{-1}^1 {\rm d}\mu P_2(\mu)\theta(\mu)
=\frac{2\theta_2}{3\pi^2}\int {\rm d}p p^4\frac{\partial f_\gamma}{\partial p}\nonumber\\
&=-\frac{8\theta_2}{3\pi^2}\int {\rm d}p p^3f_\gamma=-\frac{8}{3}\rho_\gamma \theta_2,
\end{align}
where $\theta_2$ is the quadrupole moment. We obtain for Eq.~(\ref{eq:k2})
\begin{align}
\label{eq:system7}
k^2(\phi-\psi)=-32\pi G a^2 (\rho_\gamma \theta_2+\rho_\nu \mathcal{N}_2),
\end{align}
where $\mathcal{N}_2$ is the quadrupole moment for neutrinos.

Equations (\ref{eq:system1},\ref{eq:system2},\ref{eq:system3},\ref{eq:system4},\ref{eq:system5},\ref{eq:system5bis},\ref{eq:system6},\ref{eq:system7}) gives the full evolution of the linear perturbations. 
\begin{align}
&\dot\delta+\frac{ik v}{a}-3\dot\psi=0,\\
&\dot v+H v+\frac{i k \phi}{a}=0,\\
&\dot\delta_b+\frac{ik v_b}{a}-3\dot\psi=0,\\
&\dot v_b+H v_b+\frac{ik \phi}{a}=\dot \tau R(3i\theta_1+v_b),\\
\label{eq:theta}
&\dot\theta-\dot \psi+\frac{i k \mu}{a}(\theta+\phi)=-\dot \tau (\theta_0-\theta+\mu v_b),\\
\label{eq:N}
&\mathcal{\dot N}-\dot \psi+\frac{i k \mu}{a}(\mathcal{N}+\phi)=0,\\
&3H(\dot\psi+H\phi)+\frac{k^2}{a^2}\psi=-4\pi G(\rho_{dm}\delta+\rho_b\delta_b+4\rho_\gamma\theta_0+4\rho_\nu\mathcal{N}_0),\\
&k^2(\phi-\psi)=-32\pi G a^2 (\rho_\gamma \theta_2+\rho_\nu \mathcal{N}_2).
\end{align}
The functions $\theta$ and $\mathcal{N}$ can be expanded in multipoles as seen in Eqs.~(\ref{eq:thetaL},\ref{eq:thetaL2}). For example, considering Eq.~(\ref{eq:theta}), multiplying it by $i^l\int_{-1}^1{\rm d}\mu P_l/2$ where $P_l$ are Legendre polynomials, and using $\int_{-1}^1 {\rm d}\mu P_l=2\delta_{l0}$, $\int_{-1}^1 {\rm d}\mu \mu P_l=2\delta_{l1}/3$, $P_l=[(l+1)P_{l+1}+lP_{l-1}]/(2l+1)$, we get:
\begin{align}
\dot \theta_l+\frac{k}{a(2l+1)}\Bigl[(l+1)\theta_{l+1}-l\theta_{l-1}\Bigr]-\dot\psi\delta_{l0}-\frac{k}{3a}\phi\delta_{l1}=-\dot\tau(\theta_0 \delta_{l0}-\theta_l)-\frac{i}{3}\dot\tau v_b\delta_{l1},
\end{align}
which can be written as
\begin{align}
&\dot\theta_0+\frac{k}{a}\theta_1 =\dot\psi,\\
&\dot\theta_1+\frac{k}{3a}(2\theta_2-\theta_0)-\frac{k}{3a}\phi =\dot\tau\theta_1-\frac{i}{3}\dot\tau v_b,\\
&(2l+1)\dot\theta_l+\frac{k}{a}\Bigl[(l+1)\theta_{l+1}-l\theta_{l-1}\Bigr] =(2l+1)\dot\tau \theta_l, \qquad l \geq 2.
\end{align}
The same calculation can be performed for Eq.~(\ref{eq:N}). Therefore, we can write our final system of equations, where we define $v\rightarrow iv$ and $v_b\rightarrow iv_b$ to make the velocities real:
\begin{align}
\label{eq:aaa}
&\theta'_0+k\theta_1 =\psi',\\
\label{eq:aab}
&3\theta'_1+k(2\theta_2-\theta_0)-k\phi =\tau'\Bigl(3\theta_1+v_b\Bigr),\\
\label{eq:aac}
&(2l+1)\theta'_l+k\Bigl[(l+1)\theta_{l+1}-l\theta_{l-1}\Bigr] =(2l+1)\tau' \theta_l, \qquad l \geq 2,\\
\label{eq:aad}
&\mathcal{N}'_0+k\mathcal{N}_1 =\psi',\\
\label{eq:aae}
&3\mathcal{N}'_1+k(2\mathcal{N}_2-\mathcal{N}_0)-k\phi =0,\\
\label{eq:aaf}
&(2l+1)\mathcal{N}'_l+k\Bigl[(l+1)\mathcal{N}_{l+1}-l\mathcal{N}_{l-1}\Bigr] =0, \qquad l \geq 2,\\
\label{eq:aag}
&\delta'-k v-3\psi'=0,\\
\label{eq:velocity}
&v'+\mathcal{H} v+k \phi=0,\\
\label{eq:aah}
&\delta'_b-k v_b-3\psi'=0,\\
\label{eq:aai}
&v'_b+\mathcal{H} v_b+k \phi=\tau' R(3\theta_1+v_b),\\
\label{eq:aaj}
&3\mathcal{H}(\psi'+\mathcal{H}\phi)+k^2\psi=-4\pi G a^2(\rho_{dm}\delta+\rho_b\delta_b+4\rho_\gamma\theta_0+4\rho_\nu\mathcal{N}_0),\\
\label{eq:aak}
&k^2(\phi-\psi)=-32\pi G a^2 (\rho_\gamma \theta_2+\rho_\nu \mathcal{N}_2).
\end{align}
where $\mathcal{H}=a'/a$, and as we have seen previously $R=\frac{4\rho_\gamma}{3\rho_b}$ and $\tau'=-n_e \sigma_T a$. Therefore an additional equation for $n_e$ should be included to close the system which can be done through the Peebles equation\cite{Dodelson:2003ft,Ma:1995ey} or approximately at very early times by the Saha equation.

\subsection{Initial conditions}
In order to integrate the previous equations, we need initial conditions. Taking them early enough, during the super-horizon evolution $k\eta\ll 1$, permits us to expand all functions as $X=\sum_{n=0}X^{(n)} (k\eta)^n$. In this way, we can neglect all coefficients multiplied by $k$. For example, considering the first equation, $\theta'_0+k\theta_1 =\psi'$, we get
\begin{align}
\sum_{n=0}n \theta_0^{(n)} \frac{(k\eta)^{n}}{\eta}+k\sum_{n=0}\theta_1^{(n)} (k\eta)^n-\sum_{n=0}n\psi^{(n)} \frac{(k\eta)^{n}}{\eta}=0,
\end{align}
which can be written as
\begin{align}
\sum_{n=0}n \theta_0^{(n)} (k\eta)^{n}+\sum_{n=0}\theta_1^{(n)} (k\eta)^{n+1}-\sum_{n=0}n\psi^{(n)} (k\eta)^{n}=0.
\end{align}
So, we see that $\theta_1^{(n)}$ couples to $(\theta_0^{(n+1)},\psi^{(n+1)})$, from which we can conclude that $\theta_1\sim (k\eta) \theta_0$ and hence subdominant. In the early universe, we can eliminate all coefficients multiplied by $k$. Also notice that $\tau'$ diverges at early time because the universe becomes denser and the number of interactions increase. Therefore coefficients multiplying $\tau'$ at early time should be zero, such as $3\theta_1+v_b$. Finally, we can show that for neutrinos and photons, we have $\theta_l\sim (k\eta)\theta_{l-1}$. We get
\begin{align}
&\theta'_0 =\psi',\\
\label{eq:N0}
&\mathcal{N}'_0 =\psi',\\
\label{eq:deltab}
&\delta'=\delta'_b=3\psi',\\
& 3\theta_1+v_b=0,\\
&3\theta_l=0, \qquad l \geq 2,\\
\label{eq:ICTheta}
&3\theta'_1 -k\theta_0-k\phi=0,\\
\label{eq:ICN}
&3\mathcal{N}'_1 -k\mathcal{N}_0-k\phi=0,\\
&(2l+1)\mathcal{N}'_l -kl \mathcal{N}_{l-1}=0, \qquad l \geq 2.
\end{align}
For the Eq.~$(2l+1)\mathcal{N}'_l -kl \mathcal{N}_{l-1}=0$ $(l \geq 2)$, we can consider at lowest order $\mathcal{N}_l=c_l(k\eta)^l$, from which we find $c_l=c_{l-1}/(2l+1)$. Therefore the initial condition on $\mathcal{N}_1$ provides condition for all other $\mathcal{N}_l$ $(l \geq 2)$, which can be obtained by taking derivatives of Eqs.~(\ref{eq:ICTheta},\ref{eq:ICN}), we get $\theta_1''=\mathcal{N}_1''=k(\psi'+\phi')/3$. From which we obtain $\theta_1=\mathcal{N}_1+q$ where $q$ is a function of $k$ and at most a linear function of $\eta$. It is known as the neutrino velocity isocurvature mode or relative neutrino heat flux that we will consider to be zero. Equations governing velocities are subdominant and should go as $k\eta$: from Eqs.~(\ref{eq:velocity}) and (\ref{eq:aai}), we find
\begin{align}
v=v_b=-\frac{k\eta}{2}\phi,
\end{align}
where we have used that $\mathcal{H}=1/\eta$ during radiation era.  Finally considering the gravitational equations and neglecting baryons and cold dark matter for initial conditions 
\begin{align}
\label{eq:phipsi}
&3\mathcal{H}(\psi'+\mathcal{H}\phi)=-16\pi G a^2(\rho_\gamma\theta_0+\rho_\nu\mathcal{N}_0),\\
&k^2(\phi-\psi)=-32\pi G a^2 \rho_\nu \mathcal{N}_2.
\end{align}
As we have seen previously, $\mathcal{N}_2$ can be obtained from $\mathcal{N}_1$ which can be obtained from previous initial conditions (\ref{eq:N0},\ref{eq:ICN}). Let us here for simplicity consider $\mathcal{N}_2=0$, which gives $\phi=\psi$ and differentiating Eq.~(\ref{eq:phipsi}), we get 
\begin{align}
\psi''+(5\mathcal{H}+\frac{\mathcal{H}'}{\mathcal{H}})\psi'+2(\mathcal{H}^2+\mathcal{H}')\psi=0,
\end{align}
where we have used the Friedmann equation $3\mathcal{H}^2=8\pi G a^2(\rho_\gamma+\rho_\nu)$. During the radiation era $(\mathcal{H}=1/\eta)$, we have $\psi''+4\psi'/\eta=0$ for which the solution reads as $\psi=a/\eta^3+b$. That corresponds to a fast decaying mode and a constant mode. This last mode will define our initial condition. Going back to Eq.~(\ref{eq:phipsi}) with $\psi$ constant, we get (using again Friedmann equation)
\begin{align}
\psi=-\frac{2}{\rho_\gamma+\rho_\nu} (\rho_\gamma\theta_0+\rho_\nu\mathcal{N}_0).
\end{align}
It is usually assumed that $\theta_0=\mathcal{N}_0$ and therefore 
\begin{align}
\theta_0=\mathcal{N}_0=-\frac{\psi}{2}.
\end{align}
Finally, we can integrate Eq.~(\ref{eq:deltab}) to obtain $\delta=\delta_b=3\psi$. That ends our set of initial conditions. Notice that we have never considered constants of integrations, the so-called adiabatic perturbations. If these constants are nonzero, they are called isocurvature perturbations. We summarize here the initial conditions for adiabatic perturbations 
\begin{align}
&\theta_0=-\frac{\psi}{2},\\
&\theta_1=v=v_b=-\frac{k\eta}{2}\psi,\\
&\theta_l=\mathcal{N}_l=0, \qquad l \geq 2\quad, (\text{we have neglected $\mathcal{N}_2$}),\\
&\delta=\delta_b=3\psi,\\
&\phi=\psi.
\end{align}
The initial condition for $\psi$ acts as a normalization, and can be chosen to be $\psi=1$. Integrating Eqs.~\eqref{eq:aaa}-
\eqref{eq:aak} with the previous initial conditions, we can get $\psi(\eta,k)\equiv\psi(a,k)$. The numerical solution is traditionally written as 
\begin{align}
\psi(a,k)=\frac{9}{10}\psi(a_i,k)T(k)\frac{D_1(a)}{a},
\end{align}
where $\psi(a_i,k)$ is the initial condition and can be normalized to $\psi(a_i,k)=1$, $D_1(a)$ is called the growth factor and is calculated at late time as $D_1(a)/a=\psi(a)/\psi(a_{\text{late}})$. In a flat universe without cosmological constant, we get at late time $D_1(a)=a$. In summary, from the resolution of the previous equations, we can obtain $\psi(a,k)$ or conversely $T(k)$.
\subsection{Linear power spectrum}
In the first section, we have shown how to calculate the primordial power spectrum generated during inflation, from which we can now obtain the power spectrum of matter distribution at any time. We can deduce from Eq.~(\ref{eq:aaj}) at late time 
\begin{align}
\delta= -\frac{k^2}{4\pi G a^2 \rho_{dm}} \psi=-\frac{2 a k^2}{3H_0^2\Omega_{m,0}} \psi=-\frac{3 k^2}{5H_0^2\Omega_{m,0}} \psi(a_i,k)T(k) D_1(a),
\end{align}
from which we can obtain the power spectrum for $\delta$:
\begin{align}
P(a,k)=\frac{9 k^4}{25H_0^4\Omega_{m,0}^2} T(k)^2 D_1(a)^2 P_{\psi}(k)
\end{align}
and as we have seen in the first section (Eq.~(\ref{eq:psiR}) during radiation era) $\psi(a_i,k)=2\mathcal{R}/3$, we get
\begin{align}
P(a,k)=\frac{4 k^4}{25H_0^4\Omega_{m,0}^2} T(k)^2 D_1(a)^2 P_{\mathcal{R}}(k),
\end{align}
but because $P_{\mathcal{R}}(k)=2\pi^2A_s(k/k_*)^{n_s-1}/k^3$, we have finally
\begin{align}
P(a,k)=\frac{8\pi^2 k}{25H_0^4\Omega_{m,0}^2} A_s \Bigl(\frac{k}{k_*}\Bigr)^{n_s-1}	T(k)^2 D_1(a)^2.
\end{align}
As promised, the goal of this section is achieved: we have been able to relate the initial inflation power spectrum (parametrized by $n_s$ and $A_s$) to a \textit{linear} power spectrum for the dark matter density contrast $\delta$. We will now see how to keep evolving the dark matter perturbation under the evolution of non-linear physics which occurs at small scales and later time.
\section{Non-linear corrections to the power spectrum}
\label{sec:non-lin}
While for CMB physics, the linear approximation is well-suited, structure formation is by essence a non-linear problem. No analytic solutions exist to Eqs.~\eqref{eq:system1} and \eqref{eq:system2} as soon as the non-linearities are turned on. However a great deal of innovative approaches and advanced techniques often imported from Quantum Field Theory, statistical physics or high energy physics were put foreward\cite{Ma:1995ey}. We can mention in particular, Lagrangian perturbation theory\cite{Buchert:1992ya,Bouchet:1994xp,Matsubara:2007wj,Bernardeau:2008ss,Rampf:2012xa,Valageas:2013gba}, renormalized perturbation theory\cite{Crocce:2005xy}, regularized perturbation theory (RegPT)\cite{Taruya:2012ut,Bernardeau:2015hbz}, the path integral formalism\cite{Valageas:2003gm,Valageas:2006bi}, the renormalization group flow\cite{Matarrese:2007wc,Floerchinger:2016hja}, coarse grained perturbation theory\cite{Pietroni:2008jx,Pietroni:2011iz,Anselmi:2012cn}, effective field theory\cite{Baumann:2010tm,Carrasco:2012cv} and kinetic field theory\cite{Bartelmann:2019unp}. For the purpose of this review, we will first stick to the basics: Standard Perturbation Theory (SPT) which is known to have a limited range of application, i.e. until $k \simeq 0.1\text{h }\text{Mpc}^{-1}$. In section \ref{sec:EFTofLSS}, we will show one research line to better model the small scales: the effective field theory approach.

As we have seen with Eq.~(\ref{eq:BoltzmannDM}), we can obtain different equations by taking moments of it. In this section, we are interested to study non-linearities and not relativistic corrections. In that sense, considering Newtonian physics in an expanding universe is a good approximation. Therefore, we will consider few approximations such as $\dot\psi=0$. Contrary to previous section, we will keep second order terms in the moment equations. We have for the zero moment
\begin{align}
    \dot n+\frac{1}{a}\partial_i(n v^i)+3n H=0,
\end{align}    
which is exactly Eq.~(\ref{eq:zeromoment}) where we neglected $\dot\psi$. Turning now to the first moment and defining
\begin{align}
    \int\frac{{\rm d}^3\bm{p}}{(2\pi)^3}\frac{p^2}{E^2}\hat p^i\hat p^j f=n(v^iv^j+\sigma^{ij}),
\end{align}
where $\sigma^{ij}$ is the stress tensor that we neglect, we find the non-linear first moment equation by keeping second order terms
\begin{align}
    \partial_t(n v^i)+\frac{1}{a}\partial_j(n v^i v^j)+4nHv^i+\frac{n}{a}\partial^i\phi=0.
\end{align}
Using these 2 moments equations, we find
\begin{align}
    \dot v^i+Hv^i+\frac{1}{a}v^j\partial_j v^i+\frac{1}{a}\partial_i\phi=0.
\end{align}
To close the system we need to have an equation for the gravitational potential, which from Eq.~(\ref{eq:PoissonGR}) in the non-relativistic limit gives the Poisson equation
\begin{align}
\label{eq:PoissonNewt}
\Delta\phi=4\pi G a^2 \delta\rho.
\end{align}
Defining as previously, $n=\bar {n}(1+\delta)$ with $\dot{\bar{n}}+3H\bar n=0$ and $\delta=\delta\rho/\rho$, we have
\begin{align}
&    \dot\delta+\frac{1}{a}\partial_i\Bigl[(1+\delta)v^i\Bigr]=0,\\
&\dot v^i+H v^i+\frac{1}{a}v^j\partial_j v^i+\frac{1}{a}\partial_i \phi=0,\\
&\Delta\phi=4\pi G a^2\rho \delta.
\end{align}
Defining  the velocity divergence\footnote{Notice that from this section, we will use the notation where $\theta$ defines the velocity divergence while in the previous sections it was reserved for temperature perturbations.} $\theta(t,\bm{x}) \equiv \partial_i v^i(t,\bm{x})/a(t)$ and working in Fourier space, we have from the previous equations:
\begin{align}
&\frac{\partial \delta(t,\bm{k})}{\partial t}+\theta(t,\bm{k})= - \int_{\bm{ k}_1,\bm{k}_2}\!\!\alpha(\bm{k}_1,\bm{k}_2)\theta(t,\bm{k}_1) \delta(t,\bm{k}_2), \label{eq:correc1} \\
 &\frac{\partial \theta(t,\bm{k})}{\partial t}+2 H \theta(t,\bm{k})+ \frac{3}{2}H^2\delta(t,\bm{k})= -\int_{\bm{ k}_1,\bm{k}_2}\!\!\beta(\bm{k}_1,\bm{k}_2)\theta(t,\bm{k}_1)\theta(t,\bm{k}_2), \label{eq:correc2}
\end{align} 
where the non-linear mode coupling kernels are $ \alpha(\bm{k}_1 , \bm{k}_2) = (\bm{k}_{12} \cdot \bm{k}_1)/\bm{k}_1^2$, $ \beta(\bm{k}_1 , \bm{k}_2) = \bm{k}_{12}^2 (\bm{k}_1 \cdot \bm{k}_2)/2\bm{k}_1^2 \bm{k}_2^2$, $\bm{k}_{12} = \bm{k}_1 + \bm{k}_2$.

Using Eq.~\eqref{eq:correc1} and Eq.~\eqref{eq:correc2}, one can get a second order differential equation for $\delta(t,\bm{k})$:
\begin{equation}
\label{eq:efdelta}
   \frac{ \partial^2 \delta(t,\bm{k})}{\partial t^2}+2H\frac{ \partial \delta(t,\bm{k})}{\partial t}-\frac{3}{2}H^2\delta(t,\bm{k})=S.
\end{equation}
$S$ stands for the sources of non-linearities. It reads: 
\begin{equation}
S=\int_{\bm{k}_1, \bm{k}_2}\!\!(2\pi)^3\delta(\bm{k} - \bm{k}_{12})\Big[\beta(\bm{k}_1,\bm{k}_2)\theta(t,\bm{k}_1)\theta(t,\bm{k}_2)- \frac{\partial_t}{a^2}\big(a^2\alpha(\bm{k}_1, \bm{k}_2)\theta(t,\bm{k}_1) \delta(t,\bm{k}_2)\big)\Big].    
\end{equation}
Defining an initial time $t_\ast$, the homogeneous equation can be solved using the standard Green formalism:
\begin{equation}
    \delta(t,\bm{k})=c_+(\bm{k})D_+(t)+c_-(\bm{k})D_-(t)+\int_{t_\ast}^t\frac{D_+(t')D_-(t)-D_+(t)D_-(t')}{W(t')}S(t',\bm{k}).
\label{eq:green}
\end{equation}
$c_+(\bm{k})$ and $c_-(\bm{k})$ are determined by the initial conditions. Note that in the relativistic case discussed in section \ref{sec:relat}, determining these initial condition order by order is a non-trivial step. $D_+(t)$ and $D_-(t)$ are the growing and the decaying mode solutions to equation \eqref{eq:efdelta} respectively. The Wronskian is defined as $W(t) \equiv D_+(t)\dot{D}_-(t)-\dot{D}_+(t)D_-(t)$ and for the case of matter domination $\Omega_m=1$, we have $D_+(t) \equiv D(t)=a(t) \propto t^{2/3}$ and $D_-(t)=a^{-3/2}(t)$. The case of dark energy ($\Omega_m<1$) just requires another expression for $D_{+/-}(t)$: a change in the time dependence is often enough to account for it as the small scale non-linear behavior does not in principle change much when dark energy is present on large scale.  We will discuss that in more detail in section \ref{sec:DE}.
\paragraph{} Finding a particular solution to Eq.~\eqref{eq:efdelta} is the moment when an approximation has to be made, for this review, we will describe in details SPT: it consists in assuming that the density contrast $\delta(t,\bm{k})$ and the velocity field $\theta(t,\bm{k})$ can be expanded in powers of the linear density contrast $\delta_L(t,\bm{k})$. In other words, the total density contrast can be expanded around their linear solution; the variance of the linear fluctuations is treated as a small parameter (and no vorticity is generated)
\begin{align}
      &  \delta(t,\bm{k}) = \sum_{n = 1}^\infty D^n(t)\int_{\bm{k}_1...\bm{k}_n}\!\!\!\!\!\!\!\! F_n(\bm{k}_1,\dots,\bm{k}_n) \delta_L(\bm{k}_1)\dots\delta_L(\bm{k}_n), \label{eq:SPTd}\\
      & \theta(t,\bm{k}) =-H(t)\sum_{n = 1}^\infty D^n(t)\int_{\bm{k}_1...\bm{k}_n}\!\!\!\!\!\!\!\! G_n(\bm{k}_1,\dots,\bm{k}_n)\delta_L(\bm{k}_1)\dots\delta_L(\bm{k}_n)\label{eq:SPTt}.
\end{align}
All the information about the non-linear behavior of the theory is encoded in the kernels $F_n$ and $G_n$. In some mildly non-linear range, this approximation should hold as the linear approximation holds in the early universe. However, it cannot hold for the full non-linear behavior where numerical N-body simulations have to be relied on.
\paragraph{} Plugging the Anz{\"a}tze \eqref{eq:SPTd}-\eqref{eq:SPTt}, into the left hand side of Eqs.~\eqref{eq:correc1} and \eqref{eq:correc2} and working at a given order in $\delta_L$ allows to obtain explicit (recursive) expressions for $F_n$ and $G_n$. A good exercise is to show that $F_2(\textbf{k}_1,\textbf{k}_2)=\frac{5}{7} \alpha(\textbf{k}_1,\textbf{k}_2)+\frac{2}{7}\beta(\textbf{k}_1,\textbf{k}_2)$. By doing the same manipulation, at order $n$, a general recursive expression reads \cite{Bernardeau:2001qr}:
\begin{eqnarray}
F_n(\textbf{q}_1, \ldots ,\textbf{q}_n) &=& \sum_{m=1}^{n-1} { G_m(\textbf{q}_1, \ldots ,\textbf{q}_m)
 \over{(2n+3)(n-1)}} \Bigl[(2n+1) \alpha(\textbf{k}_1,\textbf{k}_2) F_{n-m}(\textbf{q}_{m+1},
 \ldots ,\textbf{q}_n) \nonumber \\ & & +2 \beta(\textbf{k}_1, \textbf{k}_2)
 G_{n-m}(\textbf{q}_{m+1}, \ldots ,\textbf{q}_n) \Bigr] \label{Fn},
\end{eqnarray}
\begin{eqnarray}
G_n(\textbf{q}_1, \ldots ,\textbf{q}_n) &=& \sum_{m=1}^{n-1} { G_m(\textbf{q}_1, \ldots ,\textbf{q}_m)
\over{(2n+3)(n-1)}} \Bigl[3 \alpha(\textbf{k}_1,\textbf{k}_2) F_{n-m}(\textbf{q}_{m+1}, \ldots
,\textbf{q}_n) \nonumber \\ & & +2n \beta(\textbf{k}_1, \textbf{k}_2) G_{n-m}(\textbf{q}_{m+1},
\ldots ,\textbf{q}_n) \Bigr] \label{Gn}.
\end{eqnarray}
By definition of the linear regime: $F_1=G_1=1$. Such expression allows to compute kernels in principle at any order using for instance \texttt{Mathematica} \cite{Mathematica}. Some properties of the symmetrised\footnote{The kernels $F_n^{(s)}$ are symmetric with respect to the input variables.} kernels $F_n$ and $G_n$ that are worth highlighting here are:\cite{Bernardeau:2001qr}
\begin{itemize}
\item Momentum conservation implies that for $\textbf{k}=\textbf{q}_1+\cdots+\textbf{q}_n$ approaching zero, regarding less of the values of the individuals $\textbf{q}_i$, $F_n^{(s)} \propto k^2$.
\item For $p \gg q_i$,  $F_n^{(s)}(\textbf{q}_1,..,\textbf{q}_{n-2}, \textbf{p} ,-\textbf{p}) \propto k^2/p^2 $, and similarly for $G_n^{(s)}$.
\item If one of the argument $\textbf{q}_i$ of $F_n^{(s)}$ or $G_n^{(s)}$ goes to zero, there is an infrared divergence of the form $\frac{\textbf{q}_i}{q^2}$.
\end{itemize}
\paragraph{} From these non-linear kernels, it is possible to construct correlation functions that describe the non-linear regime of dark matter perturbation, as defined in Eqs.~\eqref{eq:2ptcorr}-\eqref{eq:3ptcorr}.
\subsection{Power spectrum}
\begin{figure}
\centering
\includegraphics[width=\textwidth]{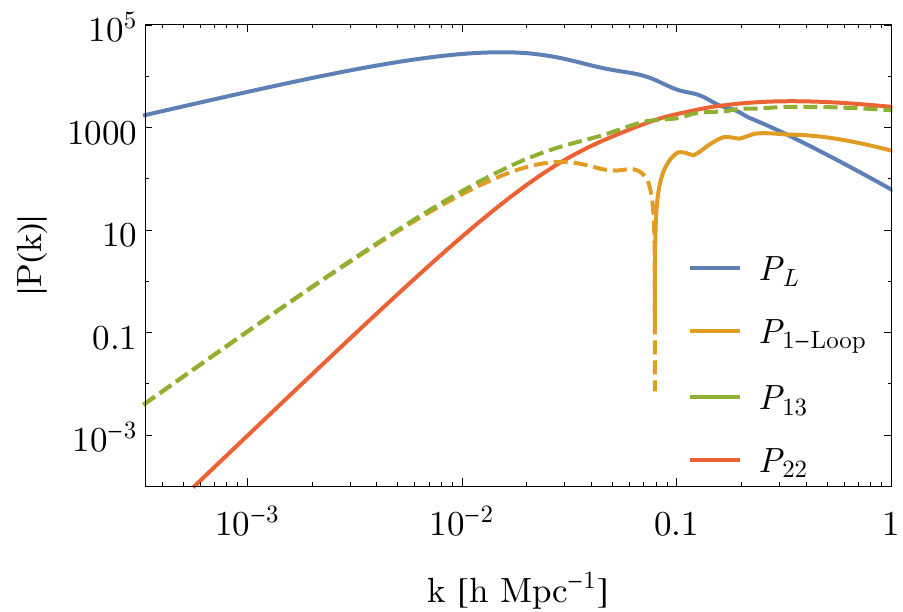}
\caption{Tree level and one-loop power spectrum. All quantities are at redshift $z=0$. Dashed lines represent a negative power spectrum. Those results agree also for instance with Ref.~\refcite{Simonovic:2017mhp}.}
\label{fig:1-loops}
\end{figure}

For the power spectrum, we can define the linear power spectrum as $P_L(k_1) \equiv \langle \delta_L(\bm{k}_1) \delta_L(\bm{k}_2)\rangle$. The first non-linear correction comes at one-loop, meaning calculating up to $n=3$. In that case, one writes:
\begin{equation}
   P_{\text{1-loop}}(t,\bm{k})=D^4(t) \left[P_{13}(\bm{k})+P_{22}(\bm{k}) \right], \label{eq:p1loop} 
\end{equation}
where two different contributions are differentiated, using Eqs.~\eqref{eq:2ptcorr} and \eqref{eq:SPTd}, they read:
\begin{align}
& P_{13}(\bm{k}) = 6P_{L}(k)\int_{\bm{q}} P_{L}(q)F_3(\bm{q},-\bm{q},\bm{k})\,, \label{eq:p13} \\
& P_{22}(\bm{k}) = 2 \int_{\bm{q}}  F_2^2(\bm{q},\bm{k}-\bm{q}) P_L(q) P_L(|\bm{k}-\bm{q}|)\,, \label{eq:p22}
\end{align}
The denomination one-loop comes from the fact that there is a loop integral with respect to the momentum $\textbf{q}$.
\paragraph{} Various techniques exist to compute the integrals \eqref{eq:p13} and \eqref{eq:p22}. See for instance, the FFT (Fast Fourier Transform) proposal in Ref.~\refcite{Simonovic:2017mhp}. Here we will present the result of a direct numerical integration using \texttt{Mathematica} \cite{Mathematica} in Fig.~\ref{fig:1-loops}.

\begin{figure}
\centering
\includegraphics[width=\textwidth]{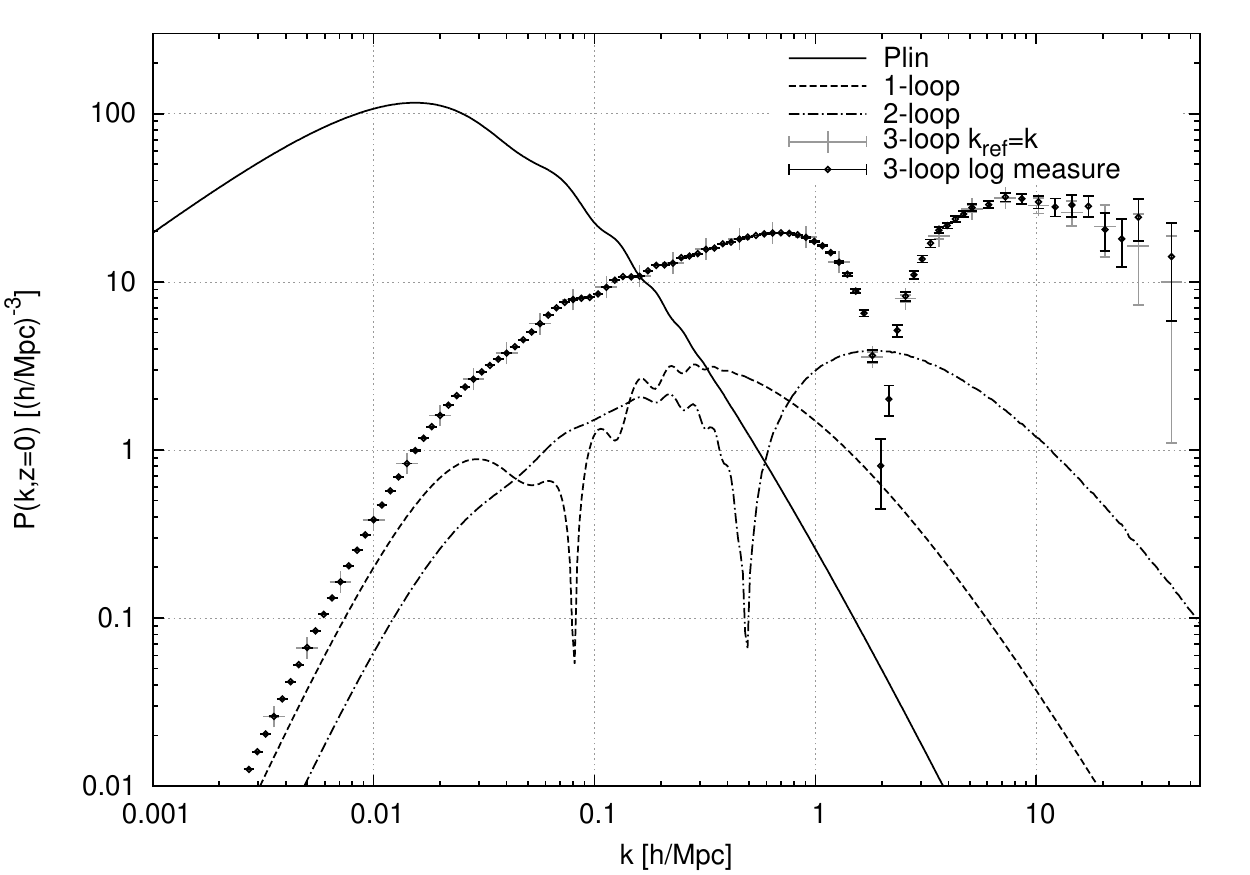}
\caption{1, 2 and 3-loop contributions to the power spectrum obtained from a numerical Monte Carlo integration within SPT at $z= 0$. Figure from Ref.~\refcite{Blas:2013aba}. Observe in particular that the 3 loop is larger than the 2 loop signalling again the poor definition of SPT in the non-linear regime.}
\label{fig:3-loop}
\end{figure}

At $k \sim 0.1 \text{h Mpc}^{-1}$ the non-linear corrections are of the same order than the linear ones and the approximation of one-loop SPT does not hold any more. In principle calculating more non-linear terms allow to further investigate the non-linear regime, but it was very early realized than brute force calculating more and more loop corrections is not a good strategy in the sense that the variance of $\delta_L$ is not small in the non-linear regime and the perturbation theory poorly converges. We will present here the equations at two-loop ($n=5$): using equations \eqref{eq:2ptcorr} and \eqref{eq:SPTd}, the different 2-loop contributions are given by 
\begin{equation}
P_{\rm 2-loop}(t,\textbf{k}) = D^6(t)\left[P_{15}(\textbf{k})+ P_{24}(\textbf{k})+P_{33}^{I}(\textbf{k}) + P_{33}^{II}(\textbf{k})   \right].
\end{equation}
With the different terms being:
\begin{align}
& P_{15} (\textbf{k}) = 30 P_{L}(k) \int_{\textbf{q}} \int_{\textbf{p}} F_5(\textbf{k},\textbf{q},-\textbf{q},\textbf{p},-\textbf{p}) P_{L}(q) P_{L}(p), \label{eq:P15} \\
& P_{24} (\textbf{k}) = 24 \int_{\textbf{q}} \int_{\textbf{p}} F_2(\textbf{q}, \textbf{k}-\textbf{q}) F_4(\textbf{p},-\textbf{p},-\textbf{q},\textbf{q}-\textbf{k}) P_{L}(q) P_{L}(p) P_{L}(|\textbf{k}-\textbf{q}|), \\
&P_{33}^{I} (\textbf{k}) = 9P_{L}(k) \int_{\textbf{q}} F_3(\textbf{k},\textbf{q},-\textbf{q}) P_{L}(q) \int_{\textbf{p}} F_3(-\textbf{k}, \textbf{p}, -\textbf{p}) P_{L}(p), \label{eq:P33I} \\
& P_{33}^{II} (\textbf{k}) = 6 \int_{\textbf{q}} \int_{\textbf{p}} F_3(\textbf{q},\textbf{p},\textbf{k}-\textbf{q}-\textbf{p}) F_3(-\textbf{q},-\textbf{p},\textbf{q}+\textbf{p}-\textbf{k}) P_{L}(q) P_{L}(p) P_{L}(|\textbf{k}-\textbf{q}-\textbf{p}|). \label{eq:P33II}
\end{align}
Those equations can be brute force integrated to obtain more precise estimations of the non-linear regime. While at one-loop, the choice of an infrared cutoff does not change much the final result, at 2-loops, it is neater to use the so-called IR-safe integrand to enjoy the IR-cancellation which occurs between the 4 contributions Eqs.~\eqref{eq:P15}-\eqref{eq:P33II}, see Ref.~\refcite{Carrasco:2013sva} where the integrals are explicitly calculated. As mentioned before, it is also possible to include some non-perturbative IR effects of the non-linear regime (resummation techniques)\cite{Matsubara:2007wj,Taruya:2012ut,Anselmi:2012cn,Senatore:2014via}.
Current calculations involve the 3-loops power spectrum, we present a result in figure \ref{fig:3-loop}.
\subsection{Bispectrum}
Most of observational surveys focus mainly on the power spectrum which is easier to measure, in particular in regard to the errors estimation (calculating a covariance matrix for a bispectrum is a difficult task). However, the non-linear bispectrum contains much more modes and information and will in principle break many degeneracies and allows to study primordial non-gaussianities.\cite{Biagetti:2019bnp} To that regards, the squeezed limit is of particular interest as we will also discuss in section \ref{sec:relat}.

The first non-zero contribution to the bispectrum comes at second order ($n=2$) in perturbation theory is:
\begin{equation}
B_{211}(k_1,k_2,k_3,t)=D^4(t)\left[ F_2(\bm{k}_1,\bm{k}_2) P_L(k_1) P_L(k_2)+\text{2 cyclic permutations}   \right].
\end{equation}
Then the 1-loop bispectrum is composed of 4 terms:
\begin{align}
& B_{\text{1-loop}}(k_1,k_2,k_3,t)=D^6(t)\left[B_{222}(k_1,k_2,k_3)+B_{321}^I(k_1,k_2,k_3) \right. \nonumber \\
& \left. +B_{321}^{II}(k_1,k_2,k_3)+B_{411}(k_1,k_2,k_3) \right],
\end{align}
using equations \eqref{eq:3ptcorr} and \eqref{eq:SPTd}, they read
\begin{align}
& B_{222}(k_1,k_2,k_3) = 8 \int_{\bm{q}} F_2 (\bm{q},\bm{k}_1-\bm{q}) F_2(\bm{k}_1-\bm{q},\bm{k}_2+\bm{q}) F_2(\bm{k}_2+\bm{q},-\bm{q})\nonumber \\
& \times P_{L}(q) P_{L}(|\bm{k}_1-\bm{q}|) P_{L}(|\bm{k}_2+\bm{q}|), \\
& B_{321}^I(k_1,k_2,k_3) = 6 P_{L}(k_1) \int_{\bm{q}} F_3(\bm{q},\bm{k}_2-\bm{q},\bm{k}_1) F_2(\bm{q},\bm{k}_2-\bm{q}) P_{L}(q) P_{L}(|\bm{k}_2-\bm{q}|) \nonumber \\ & + 5\;\text{permutations}, \\
& B_{321}^{II}(k_1,k_2,k_3) = F_2(\bm{k}_1,\bm{k}_2) P_{L}(k_1) P_{13}(\bm{k}_2) + 5\;\text{permutations}, \\
& B_{411}(k_1,k_2,k_3) = 12 P_{L}(k_1) P_{L}(k_2) \int_{\bm{q}} F_4(\bm{q},-\bm{q},-\bm{k}_1,-\bm{k}_2) P_{L}(q) \nonumber \\  &+ \text{2 cyclic permutations}.
\end{align}
We present the results of a brute force integration using the Cuba library \cite{Hahn:2004fe}. Current calculations involve the bispectrum at 3-loops\cite{Lazanu:2018yae}. 
\begin{figure}
\centering
\includegraphics[width=\textwidth]{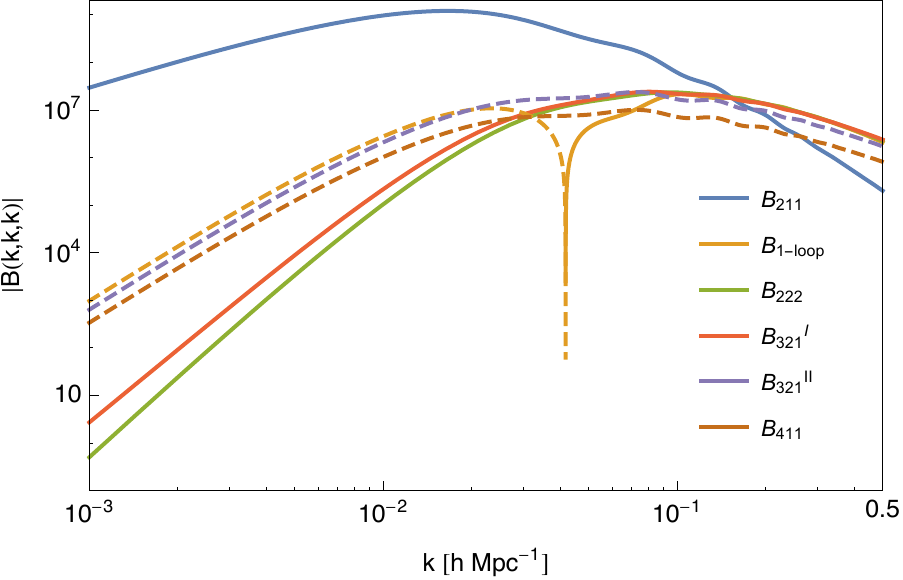}
\caption{Tree level and one-loop bispectrum in the equilateral configuration (all three momentum equal). All quantities are at redshift $z=0$. Dashed lines represent a negative bispectrum. Our results also agree with Ref.~\refcite{Baldauf:2014qfa}.}
\label{fig:1-loopsBS}
\end{figure}

\section{Relativistic corrections}
\label{sec:relat}
\paragraph{}
So far, non-linear gravitational interactions were described by Poisson equation (Eq.~\ref{eq:PoissonNewt}) in its Newtonian limit. Several research groups included some relativistic effects in their large scale structure calculations. They can be divided into three sorts: (i) relativistic corrections to the dynamics of dark matter perturbations (ii) relativistic corrections to a bias expansion relating the dark matter field to the galaxy field\cite{Yoo:2009au,Challinor:2011bk,Bonvin:2011bg,Bruni:2011ta,Baldauf:2011bh,Jeong:2011as,Yoo:2014vta,Dai:2015jaa,Desjacques:2016bnm,Fidler:2018dcy,Umeh:2019qyd} (iii) relativistic correction to the propagation of photons in a clumpy spacetime\cite{Yoo:2009au,Yoo:2010ni,Bonvin:2011bg,Jeong:2011as,Andrianomena:2014sya,Thomas:2014aga,Bertacca:2014dra,Yoo:2014sfa,DiDio:2014lka,Bonvin:2014owa,Durrer:2016jzq,DiDio:2018zmk,DiDio:2018unb} (sometimes refereed to as \textit{Redshift Space Distortion}). In this review, we will report of the first point: relativistic corrections to the dark matter dynamics. A relativistic framework is important to investigate degrees of freedom which are ignored in standard cosmology. For instance neutrinos are relativistic by definition\cite{Adamek:2017uiq}. The idea behind the modified gravity proposal for LSS is to investigate the most general modification of GR. Such modifications turn on extra degrees of freedom which have to be worked out in a relativistic framework\cite{Reverberi:2019bov}. The case for backreaction\cite{Buchert:2007ik} also requires a relativistic framework, as some backreaction terms are total derivatives in the Newtonian framework. For the case of the bispectrum which couples scales, relativistic correction are also found to be of the same order than the Newtonian results at one loop\cite{Castiblanco:2018qsd}.
.
Some groups chose to implement a relativistic N-body simulation\cite{Vogelsberger:2019ynw,Adamek:2013wja,Adamek:2015eda,Mertens:2015ttp,Bentivegna:2015flc,Adamek:2016zes,Fidler:2016tir,Daverio:2019gql,Barrera-Hinojosa:2019mzo}. The algorithms have been refined for decades and provide state of the art predictions for the non-linear regime reaching an impressive agreement with observations. Conversely, it is also important to consider analytic results which are complementary to simulations in several senses. First conceptually, one should never forget that a computer to some extend never \textit{explains} the properties of the structure but only \textit{reproduces} them with amazing details. The hope with analytic models is to be able to do fundamental physics by tracing back the features in the large scale structures from their fundamental origin. Second, analytic results are also flexible when one wants to change the assumptions of a given model, in particular when it comes to scan the parameter space of a given cosmological model. A numerical simulation being computationally expensive, running it on a high dimensional parameter space is often impossible, computationally speaking. Third, on very large scale, analytic results can often be derived while it is the range which is the harder to access as simulations are built with small scales (gravitational) interactions. In this sense analytic results are complementary to simulation. Fourth shot noise is unavoidable in simulations, as there is a finite number of particles in a given volume. Analytic results, in particular, for higher order correlation function, could help calculating such quantities without the uncertainty due to shot noise and the finite sample one can obtain from simulations.

As soon as we study non-linear relativistic effects, the scalar, vector and tensor perturbations couple to each other, which implies that they have to be considered together.
The line element of a perturbed FLRW universe can be written:

\begin{equation}
\label{eq:metric}
ds^2 = -(1+2\phi) dt^2+2 \omega_i dx^i dt + a(t)^2 \left[(1-2\psi)\delta_{ij} + h_{ij} \right]dx^i dx^j\,,
\end{equation} 
In table \ref{tab:a}, we collected some popular gauge choices where $u^{\mu}$ is the 4-velocity of the dark matter fluid.

\begin{table}
\caption{Different gauge choices for relativistic structure formation considered in this review.}
\begin{tabular*}{13cm}{@{\extracolsep{\fill}}|p{2.5cm}|>{\centering}p{2.8cm}|p{3.5cm}|p{2cm}|}
\hline 
Gauge choice & Gauge conditions & Comments & References\tabularnewline
\hline 
\hline 
Poisson or

Newtonian or

longitudinal & $\delta^{ij}\partial_{j}\omega_{i}=0$

$\delta^{ij}h_{ij}=0$

$\delta^{jk}\partial_{k}h_{ij}=0$ & Gauge chosen for the

N-body simulations

gevolution\cite{Adamek:2016zes}. Physical

interpretation is

straightforward. & \refcite{Ma:1995ey,Yoo:2009au,Bruni:2011ta,Baldauf:2011bh,Yoo:2014vta,Fidler:2018dcy,Adamek:2017uiq,Reverberi:2019bov,Andrianomena:2014sya,Thomas:2014aga,Bertacca:2014dra,DiDio:2014lka,DiDio:2018zmk,Castiblanco:2018qsd,Adamek:2013wja,Adamek:2015eda}
 \tabularnewline 
\hline 
Comoving & $\delta^{ij}h_{ij}=0$

$\delta^{jk}\partial_{k}h_{ij}=0$

$u^{0}=1$ & Easy connection with observations as time is defined by a comoving
observer. & \refcite{Yoo:2014vta,Castiblanco:2018qsd,Hwang:2015jja}\tabularnewline
\hline 
Synchronous

and comoving & $\phi=0$

$\omega_{i}=0$ & Lagrangian picture and

easy connection with observations. & \refcite{Ma:1995ey,Yoo:2009au,Challinor:2011bk,Bruni:2011ta,Jeong:2011as,Yoo:2014vta,Dai:2015jaa,Umeh:2019qyd,Bertacca:2014dra,Mertens:2015ttp,Bentivegna:2015flc} \tabularnewline
\hline 
\end{tabular*}
\label{tab:a}
\end{table}

\begin{figure}
\centering
\includegraphics[width=\textwidth]{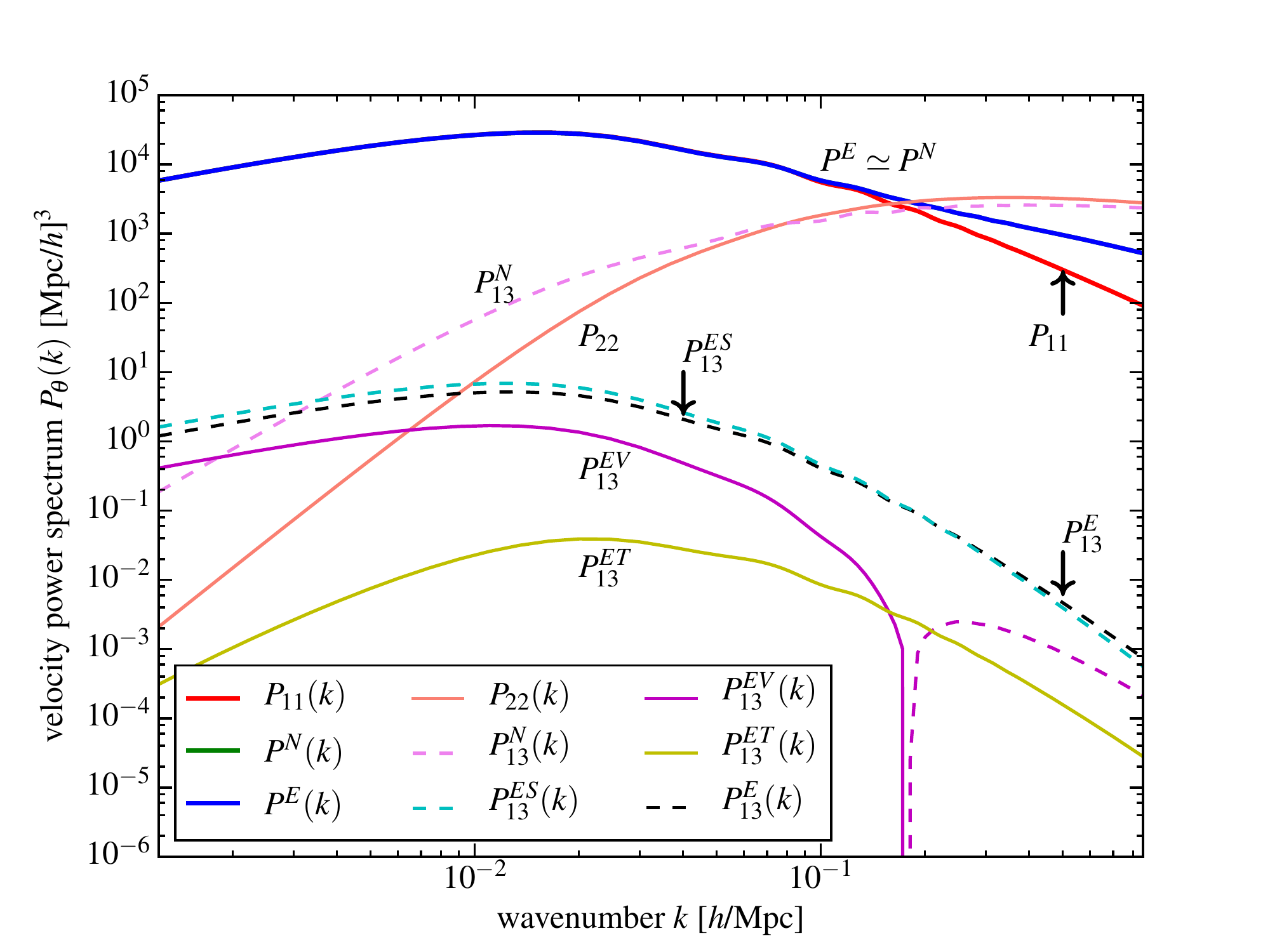}
\caption{Relativistic corrections in the comoving gauge, as adopted in Ref.~\refcite{Hwang:2015jja}. The notation are: $P_{11}\equiv P_L, P^N\equiv P_{L}+P_{\text{1-loop}}$, $P^E\equiv$ the sum of all relativistic corrections and of Newtonian results, $P^{ES}, P^{EV}, P^{ET} \equiv$ three different types of relativistic corrections corresponding to a Scalar Vector Tensor decomposition. The relativistic corrections to the matter power spectrum are 4 orders of magnitude smaller than the Newtonian result. Dashed line represent a negative power spectrum. The legend claiming to be the velocity power spectrum on the left seems to be a typo and it is the matter power spectrum.}
\label{fig:comovinggauge}
\end{figure}
GR effects to third order (n=3) in perturbation theory in the comoving gauge have been calculated in Ref.~\refcite{Hwang:2015jja}. They find agreement between GR and Newtonian physics to second order. Their result is summed up in figure \ref{fig:comovinggauge}.
Using another approach, Ref.~\refcite{Goldberg:2016lcq} calculated also relativistic corrections to the dynamics using a two-parameter expansion and identifying gauge invariant variables. 

\begin{figure}
\centering
\includegraphics[width=\textwidth]{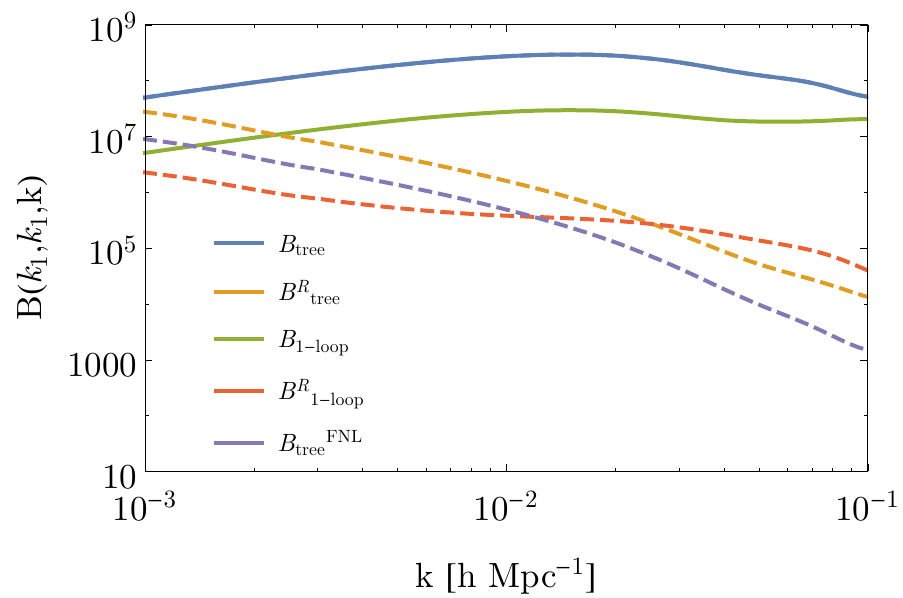}
\caption{Relativistic correction to the bispectrum in the configuration $B(k_1,k_1,k)$ where $k_1=0.1 \text{ h Mpc}^{-1}<k$. Dashed lines represent a negative bispectrum.}
\label{fig:onlbisperelcorr}
\end{figure}

Finally, the bispectrum at one-loop in a relativistic framework has been computed in Ref.~\refcite{Castiblanco:2018qsd},  
see figure \ref{fig:onlbisperelcorr}.
It was found that at one-loop the relativistic corrections are of the same order than the Newtonian results and of the same order than a primordial non-Gaussian signal with $f_{NL}=1$.
Claims that relativistic corrections matter for future (close to the horizon) surveys have to be carefully examined and then could be routinely implemented in the data analysis. 
\section{Dark energy and modified gravity} \label{sec:DE}
Even if the $\Lambda$CDM model is an extremely accurate description of the universe, the community has always searched for alternative theories. The motivations are diverse\cite{Gannouji:2019mph}. Some are constructed as a duplicate of inflationary models to dark energy models. We can consider in this category scalar field theories such as the inflaton which became quintessence or their generalized forms the so-called K-inflation \cite{ArmendarizPicon:1999rj} into K-essence \cite{ArmendarizPicon:2000dh}. The community has always been motivated to enlarge the space of models, by considering the most general model which respect some stability conditions, or some symmetries. We could name this area as a principle of no-discrimination. One includes all terms consistent with some pre-defined rules without neglecting any possible operator. These models are usually phenomenological. On the other hand, other theories are motivated by the existence of extra dimensions. These dimensions generate additional fields once studied in 4 dimensions. The Kaluza-Klein type of model gives rise to dilaton and brane models to galileon. They are often generalized and therefore become part of the first group. Finally, let us mention, maybe, a more fundamental reason. We know that the theory of general relativity is non-linear and not UV completed. Could we imagine that higher order operators in a renormalizable theory of gravity leak to the IR, and produce effects measurable at large scales? In fact, such example exists, the cosmological constant. It is generated by UV physics but it controls the expansion of the Universe, which is definitely an IR phenomenon. 
But no matter the motivation, most of these models beyond GR are in some extent phenomenological and attempt principally to see if new observables distinct to the $\Lambda$CDM model are possible. In few words, can we falsify GR from cosmological observations?

In this direction, two types of models have been studied. In the first case, one changes the background expansion of the universe by adding an additional fluid, the so-called dark energy models. In the second category, a modification of gravity is considered which translates into a modified Poisson equation and the appearance of an effective Newtonian constant which in the most general case is time and space dependent. This second category offers a very rich phenomenology. The additional fields modify the Poisson equation because of an additional force, dubbed fifth force, giving rise to an effective gravitational constant. The fifth force becomes negligible in some regime and we recover the Newtonian constant and therefore general relativity. The fifth force is screened. We can consider in this category chameleon field\cite{Khoury:2003aq,Khoury:2003rn}, which is screened in dense environment, Vainshtein mechanism\cite{Vainshtein:1972sx} which screens the extra force at small scales and finally symmetron\cite{Hinterbichler:2010es} where the field is screened also when the density of matter is small enough but because of a spontaneous symmetry breaking and a modification of the vacuum expectation value of the field.

Dark energy models and modify gravity are intrinsically not different and could be connected easily. In fact, quintessence field that is one of the most popular dark energy models, and motivated by various high energy theories, could in the absence of symmetries, couple to matter and therefore generate a scalar-tensor theory, namely a modification of gravity.

But of course, on the road to modify gravity, one should never forget the Weinberg's theorem which claims that at low energy, a Lorentz invariant theory of massless spin-2 particle must be general relativity. Therefore, any modification should either add new degrees of freedom (e.g. scalar) make the graviton massive (which via Stueckelberg mechanism also contains scalar field) or violate Lorentz invariance (such as ghost condensate).

Considering these extended theories of gravity, one can also calculate the linear and non-linear evolution of dark matter perturbation. The scheme is very similar to section 4, except that Eq.~(\ref{eq:PoissonNewt}) is replaced by a modified Poisson equation which gives instead of Eq.~(\ref{eq:efdelta})
\begin{equation}
\label{eq:DE}
   \frac{ \partial^2 \delta(t,\bm{k})}{\partial t^2}+2H\frac{ \partial \delta(t,\bm{k})}{\partial t}-\frac{3}{2}\frac{G_{\text{eff}}(t,\bm{k})}{G_N} H^2\delta(t,\bm{k})=S.
\end{equation}
where $G_{\text{eff}}$ and $G_N$ are respectively the effective and Newtonian gravitational constants. It would be impossible to do an exhaustive list of the models and their phenomenology which are built on either a modification of the second term of eq.(\ref{eq:DE}), the dark energy models or a modification of the Newtonian constant. An equation of state for dark energy $w=P/\rho<-1$ would oppose to structure formation compared to the $\Lambda$CDM while $w>-1$ would increase $\delta$. Also the effective gravitational constant would produce in modified gravity models the same phenomenology as dark energy but with an additional effect associated to scale dependence of $G_{\text{eff}}$. All scales would not structure similarly. Finally, at smaller scales or low density clusters compared to high density regions, the previously mentioned mechanisms such as Vainshtein or chameleon respectively, can generate a rich phenomenology. These models are described by additional degrees of freedom which are suppressed in some regimes. Mentioning few possible signatures of these models, we have a modification of the void lensing signal because of the presence of an additional force. Also even if it is simpler to say than to use, we can imagine to compare mass of clusters by using lensing or velocity field. In fact, lensing permits to estimate the mass of a cluster but also considering the velocity dispersion of matter falling towards the cluster, one can infer the mass. Some modified gravity models such as scalar tensor theories do not change lensing but modify the velocity dispersion by the modification of the Newtonian gravitational constant. Therefore a mismatch between these two observations could be detected. We can also mention that small galaxies for which the fifth force is not suppressed should move faster than large galaxies for which the fifth force is suppressed. The same reasons should for example produce that small galaxies stream out of voids faster than large galaxies. See e.g. Refs.~(\refcite{Jain:2007yk,Hui:2009kc,Joyce:2014kja}) for more details on possible observables.

All these new observables should be carefully calculated in the non-linear regime before testing them to observations. Even if some approximations are motivated in the standard model, they might break in modified gravity. For example, the quasi-static approximation which is wildly used, is known to be incorrect in some models\cite{Sawicki:2015zya}. In that direction, we should cite a recent paper, considering cosmological non-linear relativistic simulation within modified gravity\cite{Hassani:2019lmy}.

\section{Small scales effects: EFT of LSS} \label{sec:EFTofLSS}
As it was mentioned in section \ref{sec:non-lin}, SPT have various limitations, in the sense that it breaks down for small scales $k_{NL} \sim 0.1\text{h}/\text{Mpc}^{-1}$, where the universe becomes non-linear. There are basically three reasons for which SPT is no longer valid at small scales. One of them is that the density contrast becomes larger than unity and therefore  is not a well defined expansion parameter anymore. Second, for generic initial conditions loop integrals diverge in the ultraviolet (UV) regime, $k \geq k_{NL}$. Therefore, in order to compute the loop integrals it is necessary to use an arbitrary cutoff which makes the physical predictions to be cutoff dependent. Third, at small scales, the universe does not behave as a fluid due to shell crossing\cite{Baumann:2010tm}.

The effective field theory of large scale structure (EFTofLSS) proposed in  Refs.~\refcite{Baumann:2010tm,Carrasco:2012cv,Hertzberg:2012qn} solves these three issues by integrating out the small scales and writing the most general effective stress-energy tensor which is consistent with the symmetries  of the FLRW background, as an expansion in the relevant degrees of freedom. The ignorance about the small scales is parametrized in coefficients which characterize the effective fluid. Examples include a non-trivial state equation, the speed of sound and viscosity, they can be measured in simulations or directly from observations. 

After integrating out the small scales by doing a smoothing process on the fluid equations on a length scale\footnote{An alternative approach to the smoothing process, using path integrals  can be found in Ref.~\refcite{Carroll:2013oxa}.} $\Lambda^{-1}$, it is possible to obtain an effective long wavelength theory,  valid for  mildly nonlinear scales  $k<k_{NL}$. The relevant degrees of freedom are the long-wavelength density contrast $\delta_l$ and the velocity divergence $\theta_l=\partial_iv^i_l$. In position space, \eqref{eq:correc1} and \eqref{eq:correc2}: the continuity and Euler equations read:
\begin{align}
\frac{\partial \delta_l(t,\bm{x})}{\partial t}+\partial_i\left[(1+\delta_l(t,\bm{x}))v^i_l(t,\bm{x})\right]&=0 \ , \\
\frac{\partial\theta_l(t,\bm{x})}{\partial t} + 2H\theta_l(t,\bm{x}) + \frac{3}{2} H^2 \delta_l(t,\bm{x})&= -\frac{1}{a\bar{\rho_l}} \partial_i\partial_j\left[ \tau^{ij}\right]_\Lambda \ ,
\end{align}
where $\bar{\rho}_l$ is the long wavelength background density and $\left[\tau^{ij}\right]_\Lambda$ is the effective stress-energy tensor obtained from the smoothing. This tensor comes from the non-linear terms in the Euler equation that involves both  short and long modes, hence the effective tensor is sourced by short wavelength modes.

As $\delta$ is a stochastic field, observable quantities are obtained by computing correlation functions. They involve correlations between the effective stress-energy tensor and long modes ${\delta_l,v_l}$, that are basically couplings between the long and short modes. But, since  it is not possible to know the dynamics of the short mode, the expression for $\left[ \tau^{ij}\right]_\Lambda$ obtained from the smoothing is not very useful to compute correlations. However, long wavelength fluctuations affects the expectation value of the short modes through tidal effects, then the expectation value of the stress-energy tensor will only depend on long modes. Therefore by imposing rotational symmetry the stress energy tensor is written as an expansion in the long wavelength modes, as follow\cite{Hertzberg:2012qn}
\begin{equation}
    \tau^{ij} = \bar{\rho}_l\left[c_s^2\delta^{ij}\left(\gamma^{-1}+\delta_l\right)-\frac{c_{bv}^2}{Ha}\delta^{ij}\theta_l-\frac{3}{4}\frac{c_{sv}^2}{Ha}\left(\partial^jv^i_l +\partial^iv^j_l-\frac{2}{3}\delta^{ij}\theta_l\right)\right]+\Delta \tau^{ij}+\cdots 
\end{equation}

 $\gamma$ is the specific heat of an ordinary fluid which is used to parameterize the background pressure, $c_s$ is the speed of sound, $c_{bv}$ and $c_{sv}$ are the bulk and shear viscosity respectively, $\Delta \tau^{ij}$ is a stochastic term\footnote{Notice that $\Delta$ doesn't refer to Laplacian} that comes from the fluctuations of the short modes, and the ellipses means higher orders terms in $\delta_l$ or its derivatives. The effect of the small scales on the long scales is to induce an effective stress energy tensor with pressure and viscosity.

As $\left\{\delta_l,\theta_l,\phi_l\right\}\ll1$, SPT is used in order to compute correlation functions\footnote{Correlation functions are computed in an expansion in $\frac{k}{k_{NL}}$.}. Then, including the effective fluid terms, equation \eqref{eq:p1loop} get generalized to: 
\begin{equation}
    P_{\text{1-loop}}(t,\bm{k}) = D^4(t) \left[P_{22}(\bm{k}) + P_{13}(\bm{k}) + P_{c_{\text{comb}}^2}(\bm{k})+P_J(\bm{k})\right].
\end{equation}
$P_{22}$ and $P_{13}$ can have UV and IR divergences depending of the initial power spectrum $P_L(k)$. Then, if $P_L(k) \propto k^n$, $P_{13}(\bm{k})$ will be UV divergent for $n\geq-1$ and IR divergent for $n\leq-1$, while $P_{22}(\bm{k})$ will be UV divergent for $n\geq1/2$ and IR divergent for $n\leq-1$. Hence, $P_{\text{1-loop}}(t,\bm{k})$ is UV divergent for  $n\geq -1$ and IR divergent for $n\leq -3$.\cite{Carrasco:2012cv} 

The UV divergences of the one-loop power spectrum for a universe dominated by dark matter are obtained by expanding $P_{22}(\bm{k})$ and $P_{13}(\bm{k})$ for large $q$,
\begin{equation}
   P_{22}(\bm{k})_{q\rightarrow{\infty}} = \frac{9}{196\pi^2}k^4 \int^\Lambda dq \frac{P_L(q)^2}{q^2},
\end{equation}
\begin{equation}
    P_{13}(\bm{k})_{q\rightarrow{\infty}}=-\frac{61}{630\pi^2}k^2P_L(k)\int^\Lambda dq P_L(q).
\end{equation}
We can see that the integrals are regularized with a hard cutoff $\Lambda$, this is quite natural since the EFTofLSS was obtained by smoothing the variables over scales of order $\Lambda$. This makes the result cutoff dependent, but in this case with the addition of the new terms from $\tau_{ij}$, the final result is cutoff independent. We will see that new terms have the precise $k$ dependence to cancel the UV divergences.

The contributions from the new terms in the Euler equation are given by 
\begin{equation}
    P_{c_{\text{comb}}^2}(\bm{k}) = \langle \delta_{c_{\text{comb}}^2}(\bm{k})\delta^{(1)}(\bm{q})\rangle=-\frac{ c_{\text{comb}}^2}{9H_0^2}k^2P_L(k)\delta_D(\bm{k}+\bm{q}),
\end{equation}
where $c_{\text{comb}}^2=c_s^2+c_{bv}^2+c_{sv}^2$, and by
\begin{equation}
    P_J(\bm{k})=\langle\delta_{\Delta_J}(\bm{k})\delta^{(1)}(\bm{q})\rangle\propto k^4 P_L(k)\delta_D(\bm{k}+\bm{q}).
\end{equation}
The $k$ dependence in $P_J$ is because mass and momentum conservation imply that on large scales the corrections to the density power spectrum from these terms scale as $k^4$ rather than $k^0$. Then both $P_{c_{\text{comb}}^2}(\bm{k})$ and  $P_J(\bm{k})$ have precisely the same $k$  dependence as $P_{13}(\bm{k})$ and $P_{22}(\bm{k})$  respectively. 
Therefore, any dependence of $\Lambda$ can be absorbed by counterterms with the form $c_{\text{comb}}^2(\Lambda)$ or $\Delta^2_J(\Lambda)=\partial_i\partial_j\Delta\tau^{ij}(\Lambda)$.

In the case of the IR divergences no counterterms are needed since $P_{22}(\bm{k})$ and $P_{13}(\bm{k})$ IR divergences vanish when considering the one-loop power spectrum, this is a consequence of the equivalence principle \cite{Bernardeau:2001qr}: the IR divergences come from the convective derivatives in the equations of motion, therefore changing the reference frame changes the IR divergences, but since the correlation functions are invariant under a change of reference frame the IR divergences must cancel.

\begin{figure}
    \centering
    \includegraphics[width=\textwidth]{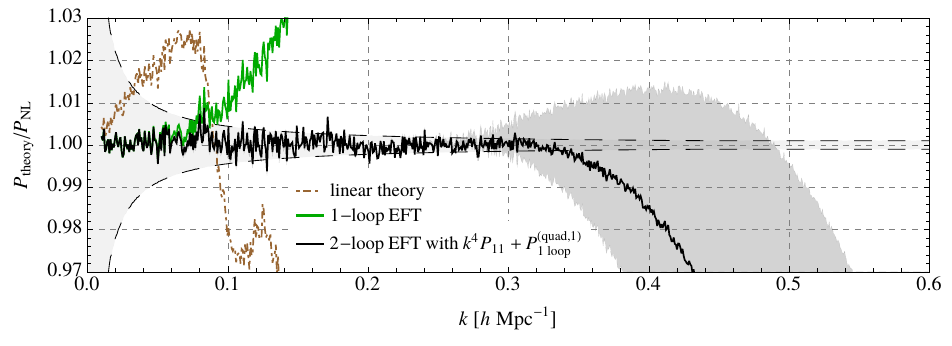}
    \caption{The prediction of the EFTofLSS at linear, one-loop and two-loop. At one-loop, only one counterterm, the speed of sound $c_s^2$, is used. At two-loops two additional counterterms, $c_1$ and $c_4$ are used. They represent a non-linear and a higher derivative speed of sound. Figure from Ref.~\refcite{Foreman:2015lca}.}
    \label{Comparison}
\end{figure}

The power spectrum up to two-loop has been also computed within the EFTofLSS \cite{Carrasco:2013mua, Carrasco:2013sva, Senatore:2014via}. The results of  Ref.~\refcite{Carrasco:2013mua} show that no new counterterm is needed beside the ones for the one-loop power spectrum at this  order  in $k/k_{NL}$. But later in Ref.~\refcite{Foreman:2015lca} a more detailed analysis is made in order to compare with N-body simulation from the Dark Sky set and they find that at two-loops three counterterms are needed, the linear term is the density contrast $\delta_l$, a quadratic term $\delta_l^2$ and a higher derivative term $\partial^2\delta_l$. The inclusion of the extra counterterms makes the predictions of the EFTofLSS  UV-insensitive and match with numerical simulations up to $k\simeq0.34\text{ h }\text{Mpc}^{-1}$ (see Fig.~\ref{Comparison} ), with better accuracy than if only a counterterm is added where the theory matches the data up to $k\sim0.15 \text{ h }\text{Mpc}^{-1}$.  

The formalism of the EFTofLSS has been extended to: (i) the two-loop power spectrum using Lagrangian perturbation theory\cite{Porto:2013qua},  (ii) the bispectrum \cite{Angulo:2014tfa, Baldauf:2014qfa}, (iii) the dark matter momentum power spectrum \cite{Senatore:2014via}, (iv) correlation functions between the density contrast and velocity including the divergence and vorticity \cite{Carrasco:2013mua, Hahn:2014lca, Mercolli:2013bsa}, (v) the baryon power spectrum \cite{Carrasco:2013mua, Lewandowski:2014rca}, (vi) the power spectrum in redshift space \cite{Senatore:2014vja}, (vii) galaxy bias \cite{Senatore:2014eva, Mirbabayi:2014zca, Angulo:2015eqa} (viii) primordial non-Gaussianities \cite{Assassi:2015jqa}, and (ix) galaxy lensing. \cite{Foreman:2015uva}.     

Very recently in Ref.~\refcite{Konstandin:2019bay} the power spectrum at three-loops has been computed, improving the level of accuracy  with respect to the two-loop results, up to $k\simeq 0.4 \text{ h } \text{Mpc}^{-1}$ at redshift $z = 0$. They also conclude that  EFTofLSS at three-loop order provides the best approximation to the power spectrum in the weakly non-linear regime at $z = 0$, and higher loops are not expected to improve the level of accuracy. 

\begin{figure}
    \centering
    \includegraphics[width=\textwidth]{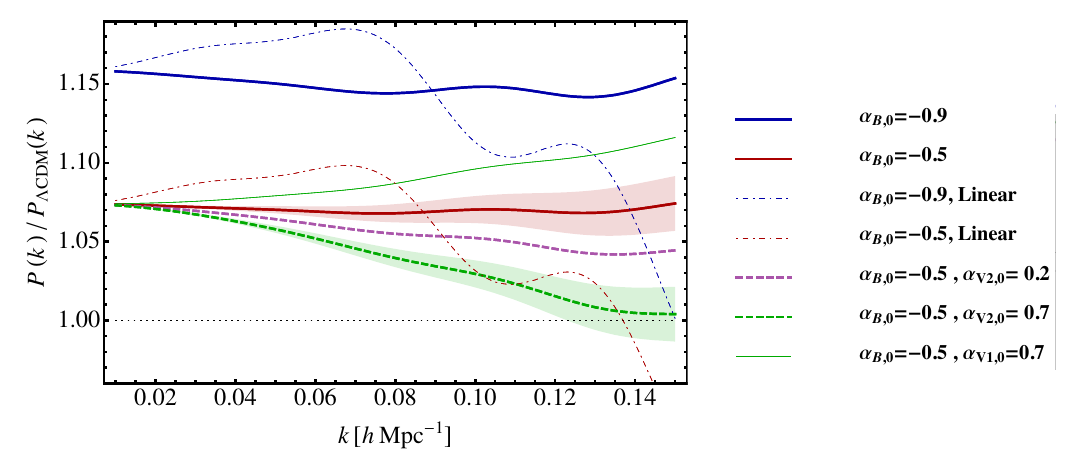}
    \caption{Effect of some of the modified gravity couplings on the one-loop matter power spectrum. Ratio of the one loop power  spectrum predicted from the  modifications of gravity and that for $\Lambda$CDM for different values of the three modified parameters $\alpha_B$, $\alpha_{V1}$ and $\alpha_{V2}$ including also the EFTofLSS counterterms.  Figure from \refcite{Cusin:2017wjg}.}
    \label{EFT+DE}
\end{figure}

Finally, in Ref.~\refcite{Cusin:2017wjg} the effects of dark energy and modified gravity on the dark matter perturbations have been studied by adding to the EFTofLSS the effective field theory of dark energy (EFTofDE) \cite{Cusin:2017mzw} in the quasi-static regime \footnote{The quasi-static approximation consists in neglecting the time evolution compared to space dynamics.}. Those results generalize the inclusion of dark energy presented in section \ref{sec:DE} by including consistently the counterterms in a framework with dark energy. As in Eq.~\eqref{eq:DE}, fluid equations conserve their form, but the inclusion of dark energy modifies the Poisson equation which is computed by solving the constraint equations of the gravitational action for Horndeski theories. Therefore, the Poisson equation is written in terms of some time dependent functions that are related to the action coefficients. Including these new functions of time in the Euler equation  and the counterterms for the EFTofLSS, the power spectrum up to one-loop is computed and the effects are showed in Fig.~\ref{EFT+DE}.

The effects caused by dark energy are parameterized in terms of the dimensionless coefficients $\alpha_B$, $\alpha_M$, $\alpha_{T}$, $\alpha_{v1}$, $\alpha_{v2}$ and $\alpha_{v3}$ (see Eq.~(39) in Ref.~\refcite{Cusin:2017wjg}). The plot shows that the non linear theory deviates from the linear theory by a few percent around $k\sim 0.1 \text{ h }\text{Mpc}^{-1}$ by the inclusion of the  coefficient $\alpha_B$ which  enters most strongly at linear level in the solution and has an relevant effect for a large change of the linear power spectrum, while $\alpha_{V1}$ and $\alpha_{V2}$ enter at cubic and higher order in the action  hence they do not enter in the lineal solution, they only modify  the power spectrum at mildly nonlinear scales, the other coefficients are set to zero. 

\section{Alternative possibility: Inhomegeneous universes:}
\label{sec:Inhom}
Cosmology is a very special field: we are actually living inside the object of study. Any cosmological observation has to be interpreted within a cosmological model. The FLRW model is the mainstream one...for good reasons: it gives a framework to interpret \textit{all} the flow of observational data collected so far in our current precision era of cosmology. However, keeping that in mind, we would like to discuss some humble works to reinterpret the observations out of the current cosmological main paradigm. Motivation may range from simply exploring the phenomenology of alternative physics, replacing the handmade addition of DE and DM fluids having unknown microscopical nature, exploring some tensions in the different data sets such as the Hubble tension\footnote{The discrepancy between the value of the Hubble constant measured with CMB and its value measured by the cosmic ladder.}.\cite{Verde:2019ivm} As examples of alternative dark matter candidates, we cite the possibility of Dirac-Milne universe\cite{BenoitLevy:2011jt,Manfredi:2018nlx} or warm dark matter.\cite{Bullock:2017xww,Ruffini:2014zfa} Now, in this section, partially adapted from chapter 6 of Ref.~\refcite{Stahl:2017oek}, we will focus on one class of model: inhomogeneous cosmologies. 

Some groups, puzzled by the \textit{ad-hoc} introduction of dark energy, tried to investigate if the relaxation of the hypothesis of spatial homogeneity could offer a decent alternative to dark energy \cite{Maartens:2011yx,Clarkson:2012bg}. Soon, it was realized that allowing for inhomogeneous spacetimes permits to fit the observed acceleration of the expansion locally\cite{Celerier:1999hp,Iguchi:2001sq}. Mathematically, the freedom that one has by introducing, say, a radial matter distribution (for a spherically symmetric universe) allows to fit any observable of the local universe such as supernovae or BAO (Baryon Acoustic Oscillations). Nonetheless, a sort of consensus has now been reached that CMB observations rule out the possibility of introducing an inhomogeneous cosmological model to explain the accelerated expansion of the universe. Indeed, in a radially inhomegeneous universe, to explain CMB observations, one needs to be placed very nearby the center of the universe which requires a huge amount of fine-tuning and disagrees with inflation\cite{Nadathur:2010zm}. 
A popular choice of inhomogeneous cosmological model is the Lema\^itre-Tolman-Bondi (LTB) metric which supposes radial inhomeneities. As FLRW, it should be understood as an approximation of the reality. Further inhomogeneous models dropping the assumption of rotational symmetry were also developed and will not be further mentioned in this review: Szekeres model\cite{Szekeres:1974ct}, Swiss cheese model \cite{Marra:2007pm}, meatball model\cite{Kainulainen:2009sx}. We will now develop a bit on the recent works involving the LTB model.

Some groups developed a fractal model within LTB metric as a way to explain the acceleration of the expansion of the universe\cite{Nogueira:2013ypa, Stahl:2016vcl,Cosmai:2018nvx}. The idea behind a self-similar pattern is to constrain the two free functions of the LTB model to a 2-parameters models that can be fitted on an equal footing with the standard $\Lambda$CDM model. For supernovae data, the results provide a decent alternative to the standard $\Lambda$CDM model. But again, these models requires some fine tuning or a transition to FLRW on larger scales as CMB observations point toward homogeneity.

A promising use of the LTB metric is in the context of modelling a light local void, but we would be situated roughly in the center of an underdense region with respect to the background FLRW metric. While for explaining the accelerated expansion of the universe, a huge void is required, the resolution of the Hubble tension\cite{Verde:2019ivm} only requires a slightly underdense region. Being on the center of such a void may have observational consequences on CMB.\cite{Nistane:2019yzd} Several claims accounting for a local void were made\cite{Clifton:2008hv,Kim:2018ihn,Lukovic:2019ryg} see however Ref.~\refcite{Kenworthy:2019qwq} for an argument against. 

\section{Concluding remarks}
This review gave an overall picture of the evolution of matter in the Universe from its origin: inflation to the moment where the Universe become structured in the non-linear regime. We presented the main equations and their physical interpretation both in the linear and non-linear regime of structures formation. We then presented some advanced topics specific to the non-linear regime of structure formation: the impact of relativistic corrections, the presence of dark energy and of small scale structures. We finally discussed an alternative view with inhomogeneous universe.
To obtain a complete picture of structures: from the big-bang to the telescope, we would need to include a link from dark matter to baryonic matter. This is usually done by introducing bias parameters. A remarkable and comprehensive review\cite{Desjacques:2016bnm} complements the references that we presented in (ii) of section \ref{sec:relat}. Finally, in cosmology, the main source of information comes from photons arriving to a telescope. These photons, are emitted by baryonic matter but also travelled through a clumpy universe, taking that effect into account is the last step to perform to obtain the full picture of structures formations in the Universe. Specialized reviews\cite{Kaiser:1987qv,Hamilton:1997zq,Percival:2008sh} can also complement the references provided in (iii) of section \ref{sec:relat}.  

A brand new front just opened recently namely the possibility to use gravitational waves for cosmology. The forecast that not only photons but also gravitons will be able to probe our universe at the largest scales delights the community. A lot of work is ahead in this direction. 
\label{sec:concl}

\section*{Acknowledgments}
We want to thank many of our collaborators, for carefull reading and comments about the manuscript, in particular Alex Gallego.
We are also very grateful to all the people who organized the Marcel Grossmann meeting: professors, staffs, secretaries, students, in particular Robert ``Bob'' Janzten.  C.S. is supported by FONDECYT Grant No 3170557, R.G. is supported by FONDECYT project No 1171384, L.C. is supported by CONICYT scholarship No 21190484.



\end{document}